\documentclass[journal,12pt,onecolumn,draftclsnofoot]{IEEEtran}
\usepackage{amsmath,amsfonts}
\usepackage{algorithmic}
\usepackage{algorithm}
\usepackage{array}
\usepackage{subcaption}
\usepackage{textcomp}
\usepackage{stfloats}
\usepackage{url}
\usepackage{verbatim}
\usepackage{color}
\usepackage{graphicx}
\usepackage[noadjust]{cite}
\usepackage{makecell}
\graphicspath{{./Figures/}}

\allowdisplaybreaks

\begin{document}

\title{Alamouti-Like Transmission Schemes \\ in Distributed MIMO Networks}

\author{Fehmi Emre Kadan,~\IEEEmembership{Member,~IEEE}, Ömer Haliloğlu,~\IEEEmembership{Member,~IEEE}, \\ and Andres Reial,~\IEEEmembership{Senior Member,~IEEE} 
\thanks{This work was supported by The Scientific and Technological Research Council of Turkey (TUBITAK) through the 1515 Frontier Research and Development Laboratories Support Program under Project 5169902, and has been partly funded by the European Commission through the Horizon Europe/JU SNS project Hexa-X-II (Grant Agreement no. 101095759). (Corresponding author: Fehmi Emre Kadan.).}
\thanks{Fehmi Emre Kadan and Ömer Haliloğlu are with Ericsson Research, Ericsson Turkey, Istanbul 34467, Turkey (e-mail: fehmi.emre.kadan@ericsson.com; omer.haliloglu@ericsson.com).}
\thanks{Andres Reial is with Ericsson Research, Ericsson AB, Lund 223 62, Sweden (e-mail: andres.reial@ericsson.com).}
}



\maketitle

\begin{abstract}
The purpose of the study is to investigate potential benefits of using Alamouti-like orthogonal space-time-frequency block codes (STFBC) in distributed multiple-input multiple-output (D-MIMO) systems to increase the diversity at the UE side when instantaneous channel state information (CSI) is not available at radio units (RUs). Most of the existing transmission techniques require instantaneous CSI to form precoders which can only be realized together with accurate and up-to-date channel knowledge. STFBC can increase the diversity at UE side without estimating the downlink channel. Under challenging channel conditions, the network can switch to a robust mode where a certain data rate is maintained for users even without knowing the channel coefficients by means of STFBC. In this study, it will be mainly focused on clustering of RUs and user equipments, where each cluster adopts a possibly different orthogonal code, so that overall spectral efficiency is optimized. Potential performance gains over known techniques that can be used when the channel is not known will be shown and performance gaps to sophisticated precoders making use of channel estimates will be identified.
\end{abstract}


\section{Introduction}
Distributed massive multiple-input multiple-output (D-MIMO, also known as cell-free massive MIMO) is a promising network type that can help to achieve extreme performance in dense urban scenarios where there would be use-cases requiring high data rates and area capacity, e.g., in public squares, stadiums, airports. Coherent Joint Transmission (CJT) is hard to realize at higher frequencies, and Non-Coherent JT (NCJT) can still provide high rates, thanks to high bandwidth, with low Spectral Efficiency (SE). Moving up in frequency, reliable communication becomes a primary concern. Macro-diversity, an implicit advantage of D-MIMO, can help us to achieve reliable access links; when it might be tricky to obtain up-to-date and accurate Channel State Information (CSI). There are practical approaches to non-coherent operation \cite{CoMPRel14}, i.e., Single Frequency Network (SFN) transmission, multi-stream transmission and even with Space-Time-Frequency Block Codes (STFBC). The aforementioned tools are designed for small number of antennas, and it is not clear how to use them with much more and distributed antennas. In traditional collocated MIMO, STFBC is an efficient technique to get the diversity from the fast fading since all the antennas are equally good on the average, but path loss and shadowing are identical. When the antennas are distributed, one can also utilize the slow fading. Although, it is still not clear how STFBC will perform in D-MIMO networks, STFBC, operating in a robust transmission mode, can provide high data rates and robustness with open-loop operation in challenging scenarios, e.g., high frequency, high mobility, lack of uplink/downlink reciprocity.

The authors in \cite{Lan03, Jin06} exploit the distributed space-time coded protocols that was devised for multi antenna systems to utilize cooperative diversity in the problems of wireless relay networks. The well-known Alamouti codes proposed in \cite{Ala98} reduces the effect of fading at the UE side by providing same diversity order as maximal ratio combining (MRC) with two-branch transmit diversity scheme. Authors in \cite{Kar19} analyzes the coverage for system information broadcast (SIB) for inactive UEs in a D-MIMO network where RUs don't have CSI information and jointly transmit a space–time block code.

Obtaining up-to-date and accurate channel state information for downlink beamforming may be tricky in distributed MIMO (D-MIMO) systems. There are several scenarios where estimating downlink channel may not be possible.

\subsection{Scenario 1 (High mobility)} 

We know that the rate of change of channel can be measured by channel coherence block length (in samples) which is defined as the multiplication of channel coherence time and channel coherence bandwidth. The channel coherence time can be viewed as the time interval in which the change in channel coefficients are small enough to assume constant (or the change in time can be well estimated by interpolation) channel coefficients. The channel coherence bandwidth is a measure of rate of change of channel through frequency, and it is determined by the multipath fading. It is inversely proportional to the multipath delay spread. For outdoor scenarios, the length of different paths are higher compared to indoor case, and hence we, in general, observe smaller coherence bandwidths for outdoor scenarios. In \cite{Ngo17}, for sub-6 GHz band in outdoor scenario with mobile users, 200 samples is assumed for channel coherence block length whereas for indoor case, this quantity becomes much larger.  

To estimate the channel with high accuracy, some pilot signals are transmitted within coherence block length. Therefore, when the coherence block length is large enough, channel estimation can be performed with a negligible error. The problems arise in the opposite case where we observe small coherence block length values. In \cite{Tor21}, it is indicated that for mm-Wave band, in outdoor scenario with a user with high mobility, we may observe very small coherence block lengths ($\sim 30$) making channel estimation hard. For time divison duplex (TDD) based transmission/reception scenario, both uplink and downlink transmissions should be performed within the coherence block making the estimation even harder. A similar problem occurs in frequency division duplex (FDD) case as the transmitter should get the user feedback within coherence block so that the channel estimates can be up-to-date. 

\subsection{Scenario 2 (Lack of Uplink/Downlink Reciprocity)} 

In TDD mode, by using the reciprocity of uplink and downlink channels, the channel estimation is performed in uplink using uplink pilots, and the estimated channel coefficients are used for both uplink payload data detection and downlink precoding. To maintain reciprocity, receive and transmit hardware chains of base stations should be calibrated. In massive MIMO framework, as the antenna elements of base stations are collocated, the calibration of receive and transmit chains can be performed with high accuracy. On the other hand, when antenna elements are geographically distributed, calibration process becomes tricky. Over the air calibration between different RUs may be required to maintain reciprocity. Under rapidly changing channel conditions, the calibration process may not be completed within the short coherence block, resulting in the lack of reciprocity. 

\subsection{Scenario 3 (Pilot Contamination)} 

To estimate the channel, some pilot signals are transmitted and it is aimed to attain orthogonal pilots for different users. When the number of users is large, it may not be possible to have different orthogonal pilot signal vectors, and one may need to reuse some pilot signals for different users. In such a case, the channel estimation errors for users associated with non-orthogonal pilots will be large and this will lead to a significant performance degradation for interference canceling methods, such as zero-forcing (ZF) \cite{Ngo13}. 

It may not be possible to have accurate and up-to-date downlink channel estimates in scenarios described above. In such a case, the network can switch to a robust transmission mode where a different transmission scheme is applied. The method which will be described in this report can be used to increase the diversity at UE side without using instantaneous channel estimates. 

\begin{figure}[ht]
\centering
  \includegraphics[width=0.75\linewidth]{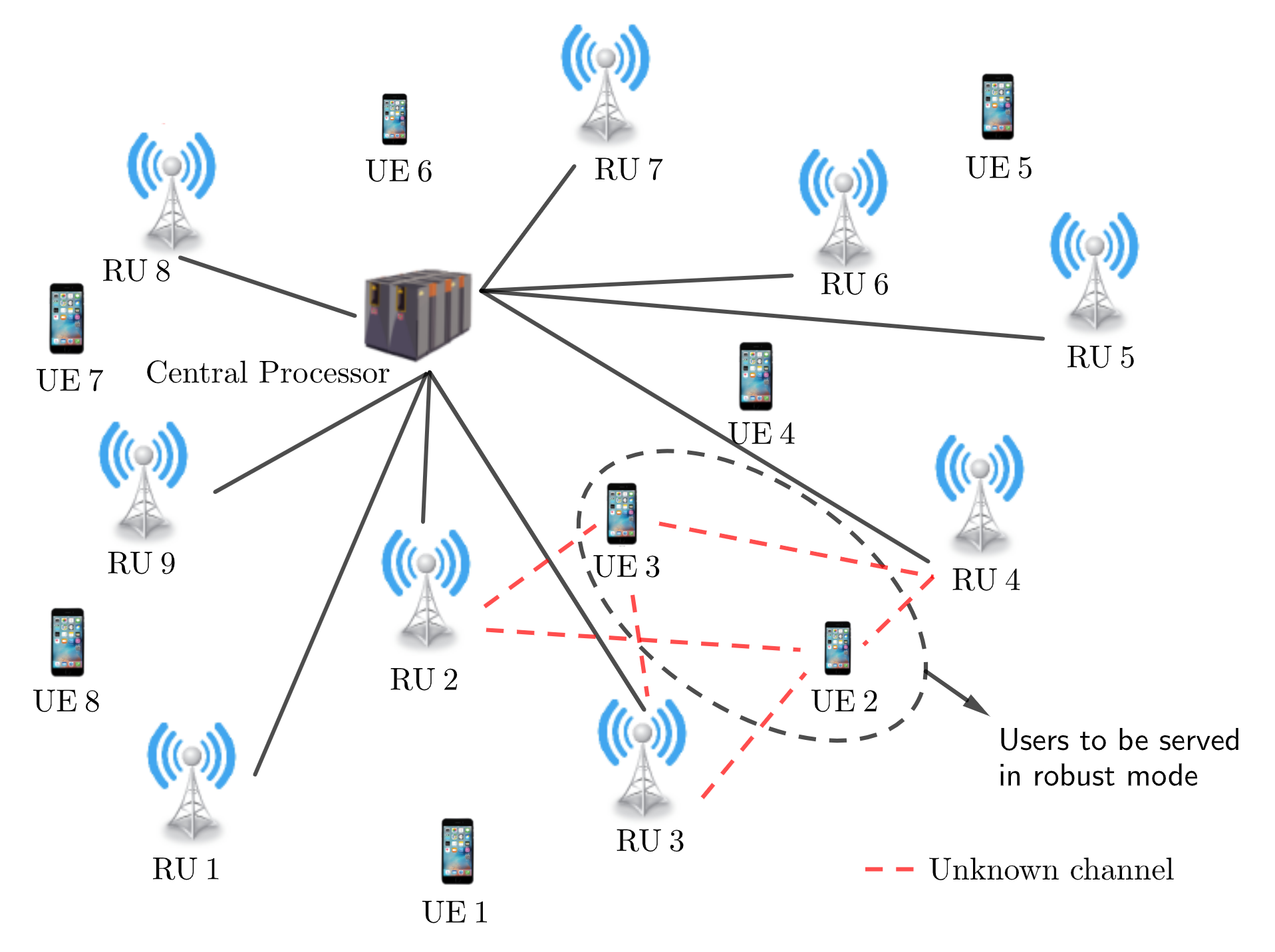}
  \caption{Robust mode for D-MIMO}
  \label{fig1}
\end{figure}

In Fig. 1, we see an example D-MIMO network with several RUs and UEs. According to the mobility and frequency band of users, and RU hardware calibration conditions, the downlink channels of some of the UEs may not be accurately estimated. The network can serve these outlying users with a robust transmission mode where downlink channel estimates are not used in precoding. In the example scenario given by Fig. 1, UE 2 and UE 3 are the ones to be served in the robust mode. The channel coefficients between UE 2-3 and RU 2-4 are not known due to a reason described by Scenarios 1-3. The channels related to other RUs cannot be estimated due to low SNR as those RUs are far away from UE 2-3. In this case, these two users can be served with robust transmission schemes that do not make use of instantaneous channel state information. 

\subsection{Organization of the Paper}

The organization of the paper is as follows. Section II includes a background related to orthogonal codes, Section III describes the system model with the usage of orthogonal coding in D-MIMO networks, Section IV involves proposed clustering method to effectively use orthogonal codes in D-MIMO, in Section V we describe other baseline methods that are considered for comparison, Section VI presents detailed simulation results and finally Section VII concludes the paper. The list of abbreviations and symbols to be used throughout the paper is given in Table I. 

\begin{table}[ht]
\centering 
\caption{List of Abbreviations and Symbols}
\begin{tabular}{|c | l |}
\hline
RU, CP, UE & radio unit, central processor, user equipment \\
\hline
CSI & channel state information \\
\hline
D-MIMO & distributed multiple-input multiple-output \\
\hline
MRT & maximal ratio transmission \\
\hline
OTFS & orthogonal time-frequency-space \\
\hline
SE & spectral efficiency \\
\hline
SINR & signal-to-interference-and-noise ratio \\
\hline
TDD, FDD & time division duplexing, frequency division duplexing \\
\hline
$s_i$ & $i$-th complex symbol transmitted to UEs \\
\hline
$h_{m,k}$ & the channel coefficient between RU $m$ and UE $k$ \\
\hline
$\textbf{h}_{k,i}$ & the channel vector of UE $k$ for symbol $s_i$ \\
\hline
$\textbf{b}_{k,i}$ & the average channel gain vector of UE $k$ for symbol $s_i$ \\
\hline
$T_k$ & the code period of orthogonal code of UE $k$ \\
\hline
$T_0$ & least common multiple of all possible code periods \\
\hline
$S$ & total number of symbols transmitted within a period $T_0$ \\
\hline
$S_k$ & the set of symbol indices transmitted to UE $k$ within a period $T_0$ \\
\hline
$C_k$ & the set of symbol indices in the cluster of UE $k$ within a period $T_0$ \\
\hline
$U_k$ & the set of RU indices serving UE $k$ \\
\hline
$V_m$ & the set of UE indices served by RU $m$ \\
\hline
$P_t$ & transmit power limit of each RU \\
\hline
$P_i$ & average transmit power of symbol $s_i$ in orthogonal coding \\
\hline
$\sigma_k^2$ & The receiver noise variance of UE $k$ \\
\hline
$\textbf{1}_{T_0}$ & all-1 vector with dimensions $T_0 \times 1$ \\
\hline
$\text{diag}(x_1, x_2, \ldots, x_K)$ & the diagonal matrix with diagonal entris $x_1, x_2, \ldots, x_K$ \\
\hline
$(\cdot)^T, (\cdot)^{*}, (\cdot)^H, \text{tr}(\cdot)$ & transpose, conjugate, conjugate-transpose, and trace operators \\
\hline
$\mathcal{C}\mathcal{N}(\mu, \sigma^2)$ & a complex Gaussian random variable with mean $\mu$ and variance $\sigma^2$ \\
\hline
$\lVert \textbf{x} \rVert$ & $\ell_2$ norm (Euclidean norm) of the vector $\textbf{x}$ \\
\hline
$[\textbf{x}]_n$ & $n$-th element of the vector $\textbf{x}$ \\
\hline
\end{tabular}ç
\label{table_1}
\end{table}

Throughout the paper, the vectors are denoted by bold lowercase letters and matrices are denoted by bold upper-case letters.

\section{Alamouti-Like Orthogonal Codes} 

In this study, to perform a robust transmission scheme, Alamouti-like orthogonal codes are used. These codes use time, frequency and space domains to increase the diversity at the receiver side. Although there are different versions of these codes (including linear combinations of transmitted symbols, non-linear codes and codes with different number fields), to make the transmission scheme simple and transmit power more uniform, we focus on transmitting only $0, s, -s, s^{*}, -s^{*}$ where $s$ is the information carrying complex symbol of a user. In other words, at any (time, frequency) grid point and for any RU antenna, only those five transformations of user complex symbols are transmitted. To make the transmission orthogonal we define a code matrix $\textbf{C}_{M, P, T}$ where $M$ denotes the number of transmit antennas, $P$ is the number of complex symbols transmitted within one code period, and $T$ stands for the code period. The matrix $\textbf{C}_{M, P, T}$ has dimensions $T \times M$ where $(i, j)$-th entry includes the transmitted signal from $j$-th antenna at the $t$-th time/frequency instant. The code is called orthogonal if the column vectors of the code matrix are mutually orthogonal for all values of transmitted complex symbols. Alamouti proposed the code with the code matrix given below \cite{Ala98}:
\begin{equation} 
\textbf{C}_{2, 2, 2} = \begin{pmatrix} s_1 & s_2 \\ -s_2^{*} & s_1^{*} \end{pmatrix}.
\end{equation} 
In this scheme, 2 complex symbols $s_1, s_2$ are transmitted within 2 time/frequency instants, from 2 transmit antennas. The code is orthogonal as the matrix $\textbf{C}_{2, 2, 2}$ is unitary for all values of $s_1$ and $s_2$.

The most important requirement for Alamouti-like orthogonal codes is that the channel should stay constant within a code period. In other words, $T \leq \tau_c$ is required where $\tau_c$ is the channel coherence block length. In Fig. 2, we see an example selection of time/frequency grid points on which the orthogonal code transmission is performed. 

\begin{figure}[ht]
\centering
  \includegraphics[width=0.75\linewidth]{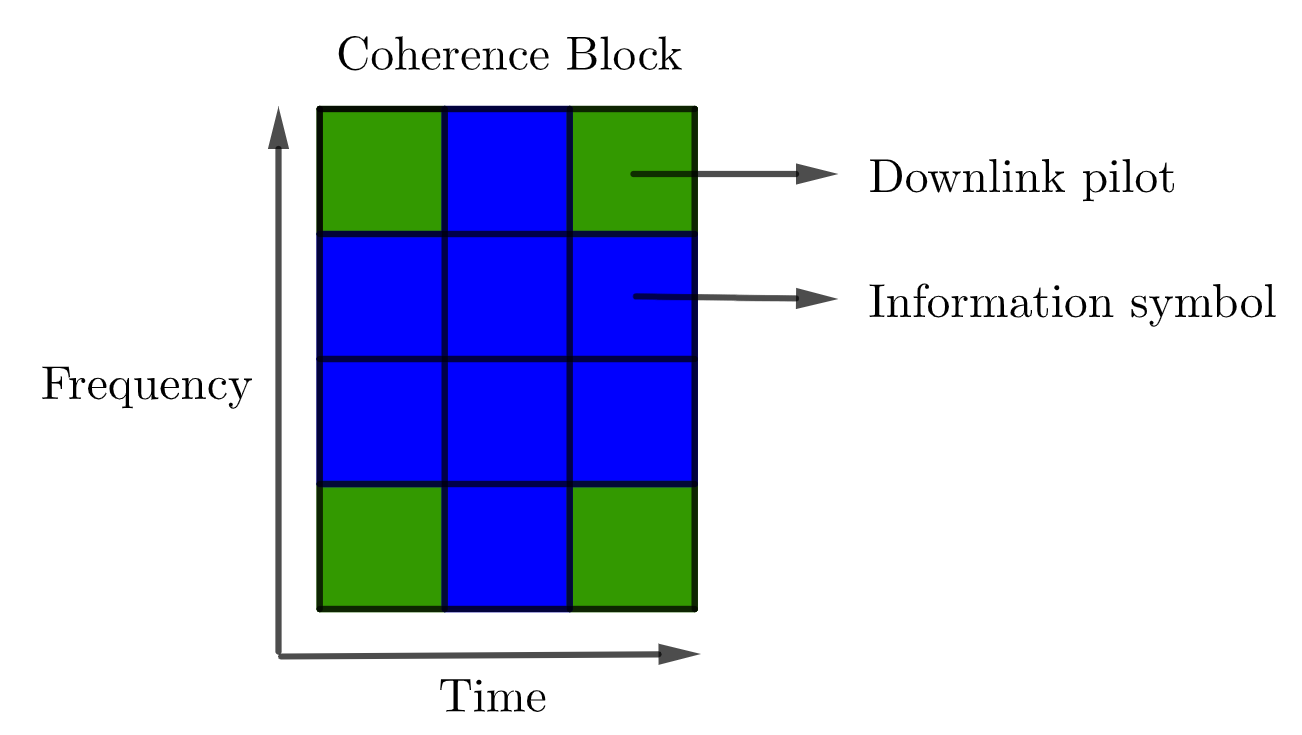}
  \caption{An example usage of time/frequency grid points}
  \label{fig2}
\end{figure}

In this example, the channel coherence block length $\tau_c=12$ and the code period $T=8$. The transmission scheme uses $8$ time/frequency grid points for information symbol transmission and other 4 remaining points are used for downlink pilot transmission. 4 mutually orthogonal pilot signals can be transmitted via this example scheme where receivers can estimate their channel coefficients for all these 12 grid points. Here, it is assumed that the channel is constant within the coherence block. 

To analyze how diversity is gained at the receiver side, we can consider $C_{2, 2, 2}$ for two different cases:

\begin{figure}[ht]
\centering
\begin{subfigure}{0.5\textwidth}
  \centering
  \includegraphics[width=0.75\linewidth]{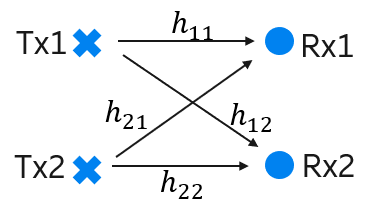}
  \caption{Single-user case}
  \label{fig3_1}
\end{subfigure}%
\begin{subfigure}{0.5\textwidth}
  \centering
  \includegraphics[width=0.75\linewidth]{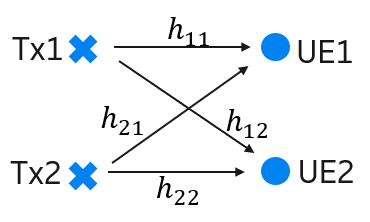}
  \caption{Multi-user case}
  \label{fig3_2}
\end{subfigure}
\caption{Single and multi-user cases}
\label{fig3}
\end{figure}

\subsection{Case 1: Single User Case} 

In single user case, $s_1, s_2$ are two complex symbols transmitted from 2 antenna elements to a UE with 2 antenna elements. Let $h_{ij}$ be the channel coefficient between $i$-th Tx antenna and  $j$-th Rx antenna. We assume that the channel stays constant throughout the code transmission (i.e., $T=2$ time/frequency instants.) Let $r_{ij}$ be the received signal by the $i$-th receive antenna at $j$-th time/frequency instant. In this case we can write
\begin{equation} 
\begin{aligned}
r_{11}&=h_{11}s_1+h_{21}s_2+n_{11} \\
r_{21}&=h_{12}s_1+h_{22}s_2+n_{12} \\
r_{12}&=h_{11}(-s_2^{*})+h_{21}s_1^{*}+n_{21} \\
r_{22}&=h_{12}(-s_2^{*})+h_{22}s_1^{*}+n_{22}
\end{aligned}
\end{equation} 
where $n_{ij}$ is the noise signal at $i$-th receive antenna at the $j$-th time/frequency instant. These four equations can be written in the matrix form as
\begin{equation} 
\underbrace{\begin{pmatrix} r_{11} \\ r_{21} \\ r_{12}^{*} \\ r_{22}^{*} \end{pmatrix}}_{\textbf{r}} = 
\underbrace{\begin{pmatrix} h_{11} & h_{21} \\ h_{12} & h_{22} \\ h_{21}^{*} & -h_{11}^{*} \\ h_{22}^{*} & -h_{12}^{*} \end{pmatrix}}_{\textbf{H}}
\underbrace{\begin{pmatrix} s_1 \\ s_2 \end{pmatrix}}_{\textbf{s}} + \underbrace{\begin{pmatrix} n_{11} \\ n_{21} \\ n_{12}^{*} \\ n_{22}^{*} \end{pmatrix}}_{\textbf{n}}.
\end{equation} 
The matrix $\textbf{H}^H\textbf{H}$ is diagonal and hence the UE can easily extract symbols $s_1, s_2$ by a simple matrix multiplication:
\begin{equation} 
\textbf{H}^H\textbf{r} = (|h_{11}|^2+|h_{12}|^2+|h_{21}|^2+|h_{22}|^2)\textbf{I}_2\textbf{s}+\textbf{H}^H\textbf{n}
\end{equation} 
where $\textbf{I}_2$ is the $2 \times 2$ identity matrix. Here it is assumed that the channel matrix $\textbf{H}$ is estimated by the receiver without any error and hence $s_1$ and $s_2$ can be perfectly separated without any inter-symbol interference. Considering the coefficient $|h_{11}|^2+|h_{12}|^2+|h_{21}|^2+|h_{22}|^2$, the UE can have 4 fold diversity using this 2 Tx, 2 Rx transmission scheme.

\subsection{Case 2: Multi User Case} 

In this case, we distribute the two receiver antennas to two different users each with a single antenna. The corresponding received signal equations can be written similarly as 
\begin{equation} 
\underbrace{\begin{pmatrix} r_{11} \\ r_{12}^{*} \end{pmatrix}}_{\textbf{r}_1} = 
\underbrace{\begin{pmatrix} h_{11} & h_{21} \\ h_{21}^{*} & -h_{11}^{*} \end{pmatrix}}_{\textbf{H}_1}
\underbrace{\begin{pmatrix} s_1 \\ s_2 \end{pmatrix}}_{\textbf{s}} + \underbrace{\begin{pmatrix} n_{11} \\ n_{12}^{*} \end{pmatrix}}_{\textbf{n}_1},
\end{equation} 
\begin{equation} 
\underbrace{\begin{pmatrix} r_{21} \\ r_{22}^{*} \end{pmatrix}}_{\textbf{r}_2} = 
\underbrace{\begin{pmatrix} h_{12} & h_{22} \\ h_{22}^{*} & -h_{12}^{*} \end{pmatrix}}_{\textbf{H}_2}
\underbrace{\begin{pmatrix} s_1 \\ s_2 \end{pmatrix}}_{\textbf{s}} + \underbrace{\begin{pmatrix} n_{21} \\ n_{22}^{*} \end{pmatrix}}_{\textbf{n}_2}.
\end{equation} 
Here $s_1$ and $s_2$ are complex symbols transmitted for UE 1 and UE 2, respectively. Using a similar operation, each UE can extract its information symbol:
\begin{equation} 
\begin{aligned}
\left[\textbf{H}_1\right]_1^H\textbf{r}_1 &= (|h_{11}|^2+|h_{21}|^2)s_1+[\textbf{H}_1]_1^H\textbf{n}_1 \\
\left[\textbf{H}_2\right]_2^H\textbf{r}_2 &= (|h_{12}|^2+|h_{22}|^2)s_2+[\textbf{H}_2]_2^H\textbf{n}_2
\end{aligned}
\end{equation} 
where $[\textbf{H}_1]_1$ is the first column of $\textbf{H}_1$ and $[\textbf{H}_2]_2$ is the second column of $\textbf{H}_2$. Notice that elements of $[\textbf{H}_1]_1$ are $h_{11}$ and $h_{21}^{*}$ and both these two quantities are related to channels of UE 1 and hence will be estimated by UE 1. Similar fact is true for UE 2 as the elements of $[\textbf{H}_2]_2$ includes channel coefficients of UE 2. Therefore, each UE can independently extract their intended complex symbols by means of the orthogonality of the code. Considering the coefficients $|h_{11}|^2+|h_{21}|^2$ and $|h_{12}|^2+|h_{22}|^2$, each UE can get 2 fold diversity considering this multi-user scheme. It is important to mention that when the multi-user scheme is applied, the transmission rate of each UE becomes halved for $(2, 2, 2)$ code as only 1 symbol is transmitted within 2 time/frequency instants for each user.

\subsection{Code Rates and Efficiency of Transmission}

In general, we can measure the efficiency of the code considering total number of complex symbols transmitted within a code period. The rate of the code is defined as the ratio $P/T$ which is equal to 1 for original (2, 2, 2) Alamouti code. In the literature, highest rate codes were found for each $M$ value \cite{Su04} for simple Alamouti-like orthogonal codes. In Table 1, the list of highest rate codes for $M \leq 8$ are given.\footnote{The table shows the highest rate code parameters when the transmission is limited to five transformations of symbols $0, s, -s, s^{*}, -s^{*}$ only. There are other linear orthogonal code schemes making use of linear combinations of different complex symbols and different fields of numbers that are not considered in this work.}
\begin{table}[ht]
\caption{Highest Rate Codes for $M \leq 8$}
\centering 
\begin{tabular}{|c | c | c | c |}
\hline
$M$ & $K$ & $T$ & Rate \\
\hline
2 & 2 & 2 & 1 \\
\hline
3 & 3 & 4 & $3/4$ \\
\hline
4 & 6 & 8 & $3/4$ \\
\hline
5 & 10 & 15 & $3/4$ \\
\hline
6 & 20 & 30 & $2/3$ \\
\hline
7 & 35 & 56 & $5/8$ \\
\hline
8 & 70 & 112 & $5/8$ \\
\hline
\end{tabular}
\label{table_2}
\end{table}
We observe in Table 1 that the rate becomes smaller as $M$ increases. Furthermore, the code period rapidly increases as $M$ increases. To make the code efficient enough and not to use a large code period to be able to perform the transmission within a coherence block, it is beneficial to use small codes. Here, the rate is measured considering all complex symbols transmitted. When multi-user transmission is performed, the rate of each UE will be less than the code rate.

To give examples for other code matrices, the ones for $M=3$ and $M=4$ are given below: 
\begin{equation} 
\textbf{C}_{3, 3, 4}=\begin{pmatrix} s_1 & s_2 & s_3 \\ -s_2^{*} & s_1^{*} & 0 \\ -s_3^{*} & 0 & s_1^{*} \\ 0 & -s_3^{*} & s_2^{*} \end{pmatrix}, \quad
\textbf{C}_{4, 6, 8}=\begin{pmatrix} s_1 & s_2 & s_3 & 0 \\ -s_2^{*} & s_1^{*} & 0 & s_4^{*} \\ -s_3^{*} & 0 & s_1^{*} & s_5^{*} \\ 0 & -s_3^{*} & s_2^{*} & s_6^{*} \\ 0 & -s_4 & -s_5 & s_1 \\ s_4 & 0 & -s_6 & s_2 \\ s_5 & s_6 & 0 & s_3 \\ -s_6^{*} & s_5^{*} & -s_4^{*} & 0 \end{pmatrix}
\end{equation}

\section{System Model} 

In this study, we assume that there are $M$ RUs each connected to a central processor (CP), and $K$ UEs. It is assumed that instantaneous channel coefficients for downlink transmission are not known, however, long-term averages of fading are known by CP and related RUs. It is also assumed that each RU and UE have multiple antennas with possibly correlated channels where channel correlation matrices are also known by CP and related RUs. Throughout the study, a centralized approach is followed where RU-UE clustering is made by CP using channel correlation matrices which are long-term quantities. To limit the fronthaul data traffic and make the transmission rate efficient, a code size limit is assumed as $T \leq 8$. Under this assumption, possible codes that can be used becomes $(M, P, T)=(2, 2, 2), (3, 3, 4), (4, 6, 8)$ considering Table 1. We can also consider a simple code without any diversity gain with parameters $(M, P, T)=(1, 1, 1)$ where single RU and single UE is involved and the user symbol is directly transmitted in a single time/frequency grid point without any transformation. This code is used in small-cell systems where each UE is served by a unique RU antenna. This code can be beneficial to serve an isolated UE with only one RU with strong channel conditions. In order for these codes to be used without significant performance degradation, the channel coherence block length value ($\tau_c$) should be larger than maximum code period $8$. The code size limit can be increased according to the channel coherence block length and the analysis that we will follow can be directly generalized. \\

\subsection{Classification of Clusters}

We can classify clusters according to the distribution of different complex symbols transmitted within a single code block among different users:

\begin{itemize}
\item \textbf{Single user, single layer:} In this type, only one user with a single complex symbol is involved in a code period. Only code that fits to this type is (1, 1, 1) code.
\item \textbf{Multi user, single layer:} In this type, multiple users each with a single complex symbol are involved in a code period. All codes with $P>1$ can be used for this type.
\item \textbf{Single user, multi layer:} In this type, multiple complex symbols are transmitted to a single user in a code period. All codes with $P>1$ can be used for this type.
\item \textbf{Multi user, multi layer:} In this type, multiple users receive possibly multiple complex symbols in a code period. All codes with $P>1$ can be used for this type.
\end{itemize}

\begin{figure}[ht]
\centering
  \includegraphics[width=\linewidth]{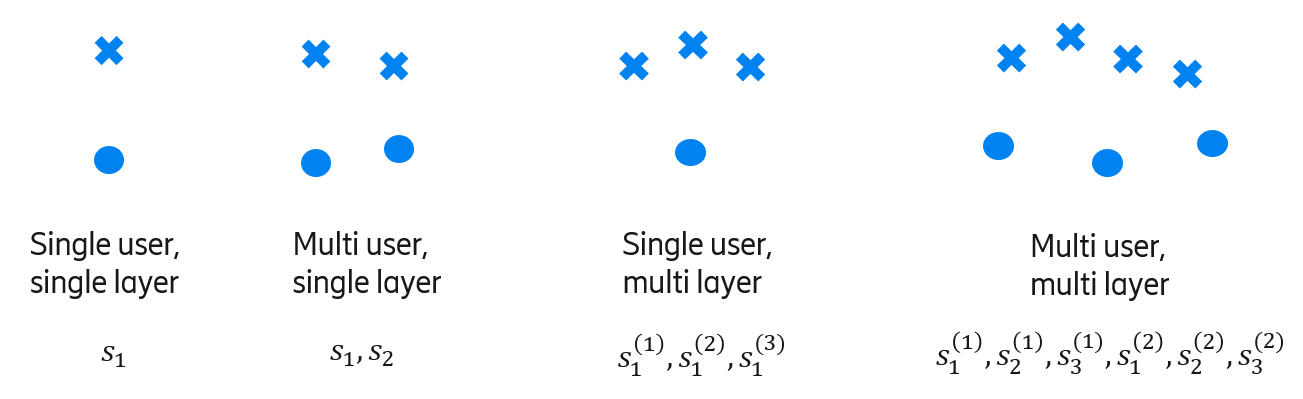}
  \caption{Different cluster types}
  \label{fig4}
\end{figure}

In Fig. 4, we present different type of clusters where transmit antennas of RUs are denoted by crosses and UEs are denoted by dots. We can also classify clusters according to the number of RUs and UEs involved. For code period limit $T \leq 8$, and the corresponding orthogonal codes $(1, 1, 1), (2, 2, 2), (3, 3, 4), (4, 6, 8)$, we can define 12 different clustering:

\begin{figure}[ht]
\centering
  \includegraphics[width=\linewidth]{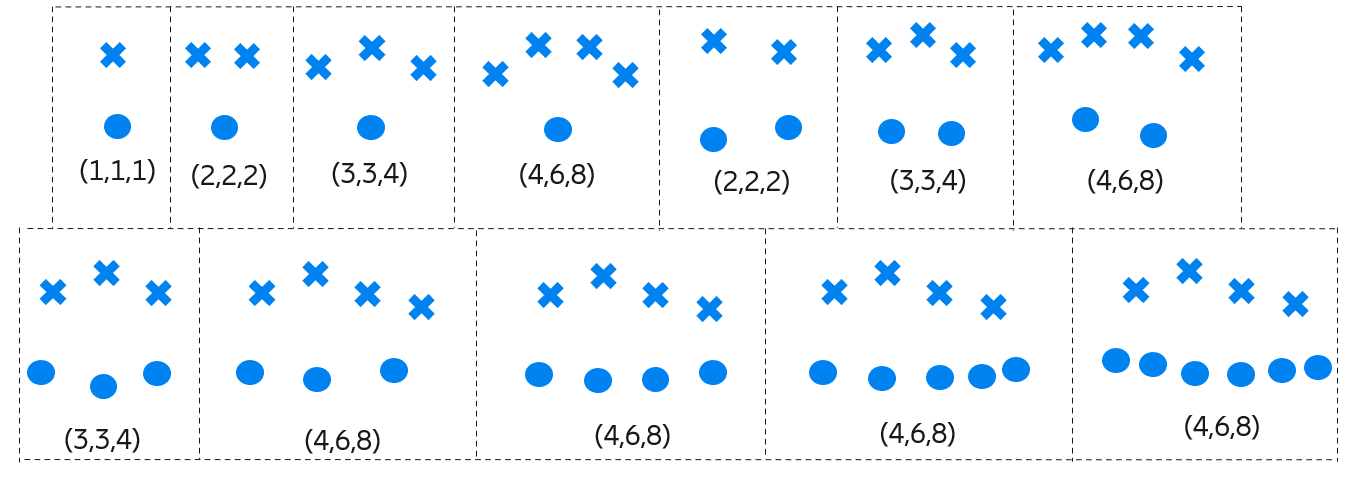}
  \caption{12 possible clusterings}
  \label{fig5}
\end{figure}

As shown in Fig. 4, at most 4 RUs and 6 UEs can be included in a cluster. Throughout the study, it is assumed that complex symbols are shared among UEs within a cluster following the ordering of UEs inside the cluster. For instance, in a cluster with RU 1, 2, 5, 6 and UE 7, 4, 5, 2, the 6 symbols to be transmitted are shared among these 4 UEs so that UE 7 and 4 receives 2 different complex symbols and UE 5 and 2 gets only one complex symbol. (Symbols are attached to UEs with indices 7, 4, 5, 2, 7, 4 following the order of UEs.) One can rearrange user ordering to obtain different symbol mappings.

Table III shows the properties of these 12 clusters:

\begin{table}[ht]
\caption{Properties of clusters}
\centering 
\begin{tabular}{|c | c | c | c | c | c |}
\hline
Cluster ID & $M$ & $K$ & $P$ & $T$ & Cluster Type \\
\hline
1 & 1 & 1 & 1 & 1 & single user, single layer \\
\hline
2 & 2 & 1 & 2 & 2 & single user, multi layer \\
\hline
3 & 3 & 1 & 3 & 4 & single user, multi layer \\
\hline
4 & 4 & 1 & 6 & 8 & single user, multi layer \\
\hline
5 & 2 & 2 & 2 & 2 & multi user, single layer \\
\hline
6 & 3 & 2 & 3 & 4 & multi user, multi layer \\
\hline
7 & 4 & 2 & 6 & 8 & multi user, multi layer \\
\hline
8 & 3 & 3 & 3 & 4 & multi user, single layer \\
\hline
9 & 4 & 3 & 6 & 8 & multi user, multi layer \\
\hline
10 & 4 & 4 & 6 & 8 & multi user, multi layer \\
\hline
11 & 4 & 5 & 6 & 8 & multi user, multi layer \\
\hline
12 & 4 & 6 & 6 & 8 & multi user, single layer \\
\hline
\end{tabular}
\label{table_3}
\end{table}

\subsection{Received Signal Equations}

To optimize and measure the performance in the network, received signal equations for orthogonal codes are derived and corresponding SINR and rate expressions are calculated. Firstly, we will write received signal expressions in terms of transformed channel and information symbol values.

To understand the main idea, we start with a simple case with (2, 2, 2) code where each RU and UE has a single antenna. In this case, the received signal vector for $k$-th user can be written as
\begin{equation}
\label{received_signal_eqn1}
\begin{pmatrix} r_{k,1} \\ r_{k,2}^{*} \end{pmatrix} = 
\begin{pmatrix} h_{m_1, k} & h_{m_2, k} \\ h_{m_2, k}^{*} & -h_{m_1, k}^{*} \end{pmatrix}
\begin{pmatrix} s_1 \\ s_2 \end{pmatrix} + \begin{pmatrix} n_{k,1} \\ n_{k,2}^{*} \end{pmatrix}
\end{equation}
where $r_{k,i}$ is the received signal sample, $n_{k,i}$ is the receiver noise sample where $i$ shows the time/frequency index for $i = 1, 2$, and $h_{m_1, k}, h_{m_2, k}$ are channel coefficient between RU $m_1$, RU $m_2$ and UE $k$, respectively. $s_1, s_2$ are two complex symbols transmitted within one code period. Notice that for single antenna case, all these quantities become complex scalars.

For (3, 3, 4) and (4, 6, 8) codes, again for single antenna case, received signal within a cluster can be written similarly as
\begin{equation}
\label{received_signal_eqn2}
\begin{pmatrix} r_{k,1} \\ r_{k,2}^{*} \\ r_{k,3}^{*} \\ r_{k,4}^{*} \end{pmatrix} = 
\begin{pmatrix} h_{m_1, k} & h_{m_2, k} & h_{m_3, k} \\ h_{m_2, k}^{*} & -h_{m_1, k}^{*} & 0 \\ h_{m_3, k}^{*} & 0 & -h_{m_1, k}^{*} \\ 0 & h_{m_3, k}^{*} & -h_{m_2, k}^{*}\end{pmatrix}
\begin{pmatrix} s_1 \\ s_2 \\ s_3 \end{pmatrix} + \begin{pmatrix} n_{k,1} \\ n_{k,2}^{*} \\ n_{k,3}^{*} \\ n_{k,4}^{*} \end{pmatrix}
\end{equation}
and
\begin{equation}
\label{r_4_6_8}
\begin{pmatrix} r_{k,1} \\ r_{k,2}^{*} \\ r_{k,3}^{*} \\ r_{k,4}^{*} \\ r_{k,5} \\ r_{k,6} \\ r_{k,7} \\ r_{k,8}^{*} \end{pmatrix} = 
\begin{pmatrix} 
h_{m_1, k} & h_{m_2, k} & h_{m_3, k} & 0 & 0 & 0 \\ 
h_{m_2, k}^{*} & -h_{m_1, k}^{*} & 0 & h_{m_4, k}^{*} & 0 & 0 \\ 
h_{m_3, k}^{*} & 0 & -h_{m_1, k}^{*} & 0 & h_{m_4, k}^{*} & 0 \\
0 & h_{m_3, k}^{*} & -h_{m_2, k}^{*} & 0 & 0 & h_{m_4, k}^{*} \\ 
h_{m_4, k} & 0 & 0 & -h_{m_2, k} & -h_{m_3, k} & 0 \\
0 & h_{m_4, k} & 0 & h_{m_1, k} & 0 & -h_{m_3, k} \\
0 & 0 & h_{m_4, k} & 0 & h_{m_1, k} & h_{m_2, k} \\
0 & 0 & 0 & -h_{m_3, k}^{*} & h_{m_2, k}^{*} & -h_{m_1, k}^{*}
\end{pmatrix}
\begin{pmatrix} s_1 \\ s_2 \\ s_3 \\ s_4 \\ s_5 \\ s_6 \end{pmatrix} + 
\begin{pmatrix} n_{k,1} \\ n_{k,2}^{*} \\ n_{k,3}^{*} \\ n_{k,4}^{*} \\ n_{k,5} \\ n_{k,6} \\ n_{k,7} \\ n_{k,8}^{*}. \end{pmatrix}
\end{equation}
UE $k$ may receive a subset of symbols transmitted in a code period according to the cluster type. According to the symbols belonging to UE $k$, the received signal of UE $k$ can be written in matrix form as
\begin{equation}
\label{r_k}
\textbf{r}_k = \underbrace{\displaystyle\sum_{i \in S_k} \textbf{h}_{k,i}s_i}_{\text{desired}} + \underbrace{\displaystyle\sum_{i \not\in S_k} \textbf{h}_{k,i}s_i}_{\text{interference}} + \underbrace{\textbf{n}_k}_{\text{noise}}
\end{equation}
where $\textbf{r}_k$ is the received signal vector of UE $k$ consisting of received signal samples within the code period of the orthogonal code that UE $k$ is involved, $\textbf{h}_{k,i}$ is the channel vector corresponding to the symbol $s_i$, $S_k$ is the set of symbol indices that UE $k$ receives, and $\textbf{n}_k$ is the receiver noise vector obtained in a code period. The channel vector $\textbf{h}_{k,i}$ is determined by cluster type and the order of UE $k$ in that cluster. For instance, if UE $k$ is involved in a cluster with RU 1, 2, 3, 4 and UE 1, $k$, 2, where each UE receives 2 complex symbols out of 6, the channel vectors for UE $k$ symbols becomes
\begin{equation}
\label{example_h_matrix_eqn}
\begin{pmatrix} 
h_{2, k} \\ -h_{1, k}^{*} \\ 0 \\ h_{3, k}^{*} \\ 0 \\ h_{4, k} \\ 0 \\ 0
\end{pmatrix}
\quad \text{and} \quad
\begin{pmatrix} 
0 \\ 0 \\ h_{4, k}^{*} \\ 0 \\ -h_{3, k} \\ 0 \\ h_{1, k} \\ h_{2, k}^{*}
\end{pmatrix}. 
\end{equation}
UE $k$ is at the second order in the UE list with 3 users and hence second and fifth columns of the matrix given in (\ref{r_4_6_8}) are taken.

In the interference part, there may be clusters with code period different from the code period of UE $k$. We assume that all orthogonal codes are time synchronized so that each code with period $T_k$ is started at time instants $t=\ell T_k$ where $\ell$ is a non-negative integer and $T_k$ is the code period for UE $k$. To calculate interference part, we consider the least common multiple of all possible code periods which is equal to $T_0=8$ since $T_k \in \{1, 2, 4, 8\}$ for all users $k$. We consider $T_0$ time/frequency samples in each cluster to measure desired and interference signal parts. For code periods less than $T_0$, we consider multiple periods of the orthogonal code in which independent symbols are transmitted.

To better understand the desired and interference part in (\ref{r_k}), we consider an example clustering scheme. In this scheme we have RU 1, 2, 3, 4, 5, 6 and UE 1, 2, 3, 4 and clusters are formed as $C_1 = \{\text{RU} 1, \text{RU} 2, \text{RU} 3, \text{RU} 4, \text{UE} 1, \text{UE} 2\}$ and $C_2 = \{\text{RU} 5, \text{RU} 6, \text{RU} 3, \text{RU} 4\}$. In this example scenario, the received signal for UE 2 can be written as 
\begin{equation}
\label{example_eqn1}
\begin{aligned}
&\textbf{r}_2 = 
\underbrace{\begin{pmatrix} 
h_{2, 2} & 0 & 0 \\ -h_{1, 2}^{*} & h_{4, 2}^{*}  & 0 \\ 0 & 0 & 0 \\ h_{3, 2}^{*} & 0 & h_{4, 2}^{*} \\ 
0 & -h_{2, 2} & 0 \\ h_{4, 2} &  h_{1, 2} & -h_{3, 2} \\ 0 & 0 & h_{2, 2} \\ 0 & -h_{3, 2}^{*} & -h_{1, 2}^{*}
\end{pmatrix} 
\begin{pmatrix} s_2 \\ s_4 \\ s_6 \end{pmatrix}}_{\text{desired}} 
+
\underbrace{\begin{pmatrix} 
h_{1, 2} & h_{3, 2} & 0 \\ h_{2, 2}^{*} & 0 & 0 \\ h_{3, 2}^{*} & -h_{1,2}^{*} & h_{4,2}^{*} \\ 0 & -h_{2,2}^{*} & 0 \\ 
h_{4, 2} & 0 & -h_{3,2} \\ 0 & 0 & 0 \\ 0 & h_{4,2} & h_{1,2}\\ 0 & 0 & h_{2,2}^{*} 
\end{pmatrix}
\begin{pmatrix} s_1 \\ s_3 \\ s_5 \end{pmatrix}}_{\text{intra-cluster interference}}
+\\
&\underbrace{\begin{pmatrix} 
h_{5, 2} & 0 & 0 & 0 \\ h_{6, 2}^{*} & 0 & 0 & 0 \\ 
0 & h_{5, 2} & 0 & 0 \\ 0 & h_{6, 2}^{*} & 0 & 0 \\
0 & 0 & h_{5, 2} & 0 \\ 0 & 0 & h_{6, 2}^{*} & 0 \\
0 & 0 & 0 & h_{5, 2} \\ 0 & 0 & 0 & h_{6, 2}^{*} \\
\end{pmatrix} 
\begin{pmatrix}
s_7 \\ s_9 \\ s_{11} \\ s_{13}
\end{pmatrix} 
+
\begin{pmatrix} 
h_{6, 2} & 0 & 0 & 0 \\ -h_{5, 2}^{*} & 0 & 0 & 0 \\ 
0 & h_{6, 2} & 0 & 0 \\ 0 & -h_{5, 2}^{*} & 0 & 0 \\
0 & 0 & h_{6, 2} & 0 \\ 0 & 0 & -h_{5, 2}^{*} & 0 \\
0 & 0 & 0 & h_{6, 2} \\ 0 & 0 & 0 & -h_{5, 2}^{*} \\
\end{pmatrix} 
\begin{pmatrix}
s_8 \\ s_{10} \\ s_{12} \\ s_{14}
\end{pmatrix}}_{\text{inter-cluster interference}}
+
\underbrace{\begin{pmatrix} n_{2,1} \\ n_{2,2}^{*} \\ n_{2,3}^{*} \\ n_{2,4}^{*} \\ n_{2,5} \\ n_{2,6} \\ n_{2,7} \\ n_{2,8}^{*} \end{pmatrix}}_{\text{noise}}.
\end{aligned}
\end{equation}
In this example, the symbols $s_2, s_4, s_6$ belong to UE 2 and hence the signal part including $s_2, s_4, s_6$ is the desired part. The symbols $s_1, s_3, s_5$ are transmitted to UE 1 which is also involved in the same cluster as UE 2. Hence, the signal part related to $s_1, s_3, s_5$ can be treated as intra-cluster interference. Notice that by the orthogonality of the code, this interference part can be eliminated by a simple matrix multiplication operation. The symbols $s_7, s_8, \ldots, s_{14}$ belong to users 3 and 4 which are involved in other cluster and hence the related signal part is referred to as inter-cluster interference. These two users are served by (2, 2, 2) code having code period 2. As the code period of cluster covering UE 2 is 8, we consider 4 periods of (2, 2, 2) code to analyze the inter-cluster interference part. The symbols $s_7, s_9, s_{11}, s_{13}$ belong to UE 3 and $s_8, s_{10}, s_{12}, s_{14}$ are transmitted for UE 4 within this 8 time/frequency instants.

We can similarly write received signal for UE 4 as
\begin{equation}
\label{example_eqn2}
\begin{aligned}
\textbf{r}_4 &= \underbrace{\begin{pmatrix} 
h_{6, 4} & 0 & 0 & 0 \\ -h_{5, 4}^{*} & 0 & 0 & 0 \\ 
0 & h_{6, 4} & 0 & 0 \\ 0 & -h_{5, 4}^{*} & 0 & 0 \\
0 & 0 & h_{6, 4} & 0 \\ 0 & 0 & -h_{5, 4}^{*} & 0 \\
0 & 0 & 0 & h_{6, 4} \\ 0 & 0 & 0 & -h_{5, 4}^{*} \\
\end{pmatrix}
\begin{pmatrix}
s_8 \\ s_{10} \\ s_{12} \\ s_{14}
\end{pmatrix}}_{\text{desired}} 
+
\underbrace{\begin{pmatrix} 
h_{5, 4} & 0 & 0 & 0 \\ h_{6, 4}^{*} & 0 & 0 & 0 \\ 
0 & h_{5, 4} & 0 & 0 \\ 0 & h_{6, 4}^{*} & 0 & 0 \\
0 & 0 & h_{5, 4} & 0 \\ 0 & 0 & h_{6, 4}^{*} & 0 \\
0 & 0 & 0 & h_{5, 4} \\ 0 & 0 & 0 & h_{6, 4}^{*} \\
\end{pmatrix} 
\begin{pmatrix}
s_7 \\ s_9 \\ s_{11} \\ s_{13}
\end{pmatrix}}_{\text{intra-cluster interference}} \\
&+ 
\underbrace{\begin{pmatrix} 
h_{1, 4} & h_{2, 4} & h_{3, 4} & 0 & 0 & 0 \\ 
h_{2, 4}^{*} & -h_{1, 4}^{*} & 0 & h_{4, 4}^{*} & 0 & 0 \\ 
h_{3, 4}^{*} & 0 & -h_{1, 4}^{*} & 0 & h_{4, 4}^{*} & 0 \\
0 & h_{3, 4}^{*} & -h_{2, 4}^{*} & 0 & 0 & h_{4, 4}^{*} \\ 
h_{4, 4} & 0 & 0 & -h_{2, 4} & -h_{3, 4} & 0 \\
0 & h_{4, 4} & 0 & h_{1, 4} & 0 & -h_{3, 4} \\
0 & 0 & h_{4, 4} & 0 & h_{1, 4} & h_{2, 4} \\
0 & 0 & 0 & -h_{3, 4}^{*} & h_{2, 4}^{*} & -h_{1, 4}^{*}
\end{pmatrix} 
\begin{pmatrix} s_1 \\ s_2 \\ s_3 \\ s_4 \\ s_5 \\ s_6 \end{pmatrix}}_{\text{inter-cluster interference}} 
+
\underbrace{\begin{pmatrix} n_{4,1} \\ n_{4,2}^{*} \\ n_{4,3} \\ n_{4,4}^{*} \\ n_{4,5} \\ n_{4,6}^{*} \\ n_{4,7} \\ n_{4,8}^{*} \end{pmatrix}}_{\text{noise}}
\end{aligned}
\end{equation}
where in this case $s_8, s_{10}, s_{12}, s_{14}$ are desired symbols for UE 4, $s_7, s_9, s_{11}, s_{13}$ are symbols of UE 3 involved in the same cluster as UE 4, and $s_1, s_2, \ldots, s_6$ are symbols transmitted in Cluster 1 to UE 1, 2.

\subsection{Detection at UE side}

In the previous section, we explained how to determine desired and interference signal components according to the cluster types, user orders within the clusters, and code periods of clusters. To make the further analysis simpler, we will continue with the general equation given in (\ref{r_k}). At user side, by means of downlink pilots, the channel estimation should be performed prior to data extraction. We assume perfect channel estimation at UE side to measure the system performance. The results to be found can be treated as upper bounds for performance evaluation.

From now on, we will consider the received signal model for multi-antenna case. We define augmented channel vector $\textbf{h}_{k,n,i}$ as the channel for $n$-th antenna of the $k$-th UE for symbol $s_i$. This vector includes channels of serving RUs according to the cluster type of UE $k$ and the order of symbol $s_i$ in that cluster. The received signal by the $n$-th antenna of UE $k$ can be written as 
\begin{equation}
\label{r_k2}
\textbf{r}_{k,n} = \displaystyle\sum_{i \in S_k} \textbf{h}_{k,n,i}s_i + \displaystyle\sum_{i \in C_k\setminus S_k} \textbf{h}_{k,n,i}s_i + \displaystyle\sum_{i \not\in C_k} \textbf{h}_{k,n,i}s_i + \textbf{z}_{k,n}
\end{equation}
where $S_k$ is the index set of symbols intended for UE $k$, $C_k$ is the index set of symbols transmitted in the cluster of UE $k$, $\textbf{h}_{k,n,i}$ is the channel vector for the $i$-th symbol and $n$-th antenna of UE $k$, $\textbf{z}_{k,n} \sim \mathcal{C}\mathcal{N}(\textbf{0}, \sigma_k^2\textbf{I})$ is the noise vector at $n$-th antenna of UE $k$. 

In (\ref{r_k2}), we consider all symbols transmitted within $T_0$ time/frequency instants. Thanks to the orthogonality of the code ($\textbf{h}_{k,n,i}^H\textbf{h}_{k,n,j}=0$ for $i \neq j \in C_k$) and disjoint nature of clusters, the symbols in $C_k \setminus S_k$ can be eliminated by the $k$-th UE whereas the symbols outside the cluster $C_k$ cannot be eliminated. Therefore, we can define an SINR term for the symbol $s_i$ for any $i \in S_k$ as
\begin{equation}
\label{SINR_eqn}
\text{SINR}_{k,i} = \dfrac{\displaystyle\sum_{n=1}^{N_k} \textbf{h}_{k,n,i}^H\textbf{h}_{k,n,i}P_t}{\displaystyle\sum_{j \not\in C_k} \mathbb{E}\left[\displaystyle\sum_{n=1}^{N_k}\textbf{h}_{k,n,j}^H\textbf{h}_{k,n,j}\right]P_t+\sigma_k^2}.
\end{equation} 
Here $N_k$ is the total number of antennas at UE $k$ and we treat the interference and noise terms as unknowns with known statistics. On the other hand, the term $\textbf{h}_{k,n,i}$ is assumed to be known by the $k$-th UE. In this study, we assume that there is a separate power amplifier for each transmit antenna at RU side, and hence a transmit power limit per antenna is assumed.

\subsection{Achievable User Rates}

As the instantaneous channel is not known at RU side, the instantaneous SINRs are random and hence to find achievable user rates, we can follow different approaches. Under fast fading assumption where the channel coding period includes all possible channel states, we can evaluate the ergodic capacity by evaluating the mean value of the Shannon capacity, i.e.,
\begin{equation}
\label{SE_ergodic_eqn}
\text{SE}_{\text{ergodic}, k, i}=\mathbb{E}[\log_2(1+\text{SINR}_{k,i})].
\end{equation}
If the joint pdf of channel vectors is known, we can evaluate the ergodic capacities of users. Another approach is to consider outage capacity where we assume that the signal can be decoded without any error if the instantaneous SINR is larger than some threshold. Assuming that no correct decoding can be done below threshold value $\text{SINR}_{\text{min}, k, i}$, we can measure outage capacity as
\begin{equation}
\label{SE_outage_eqn}
\text{SE}_{\text{outage}, k, i}=(1-P_{\text{out}})\log_2(1+\text{SINR}_{\text{min}, k, i})
\end{equation}
where
\begin{equation}
\label{SE_outage_eqn2}
\text{Pr}(\text{SINR}_{k,i}<\text{SINR}_{\text{min}, k, i})=P_{\text{out}}.
\end{equation}
In general, the outage capacity is used when we observe slow fading and its calculation requires joint pdf of channel vectors. A general approach is to set a $P_{\text{out}}$ and evaluate corresponding $\text{SINR}_{\text{min}, k, i}$ and $\text{SE}_{\text{outage}, k, i}$ values. To avoid frequent retransmissions needed for symbols in outage, a sufficiently small $P_{\text{out}}$ is needed.

In this study, we will consider both these rate expressions in cluster formation and performance comparison. 

\subsection{Assumptions About Channel}

In this study, we assume that second order long-term channel statistics are known by CP and related RUs for all RU-UE pairs. We also assume Rayleigh small scale fading where all small-scale fading coefficients has the distribution $\mathcal{C}\mathcal{N}(0, 1)$. The channels corresponding to different RUs and different UEs are independent, but the channels of the antennas of the same RU and similarly the channels of the antennas of the same UE are correlated.

\subsection{Evaluation of Spectral Efficiencies}

Using the assumptions about channel coefficients, we can evaluate the pdf of $\text{SINR}_{k,i}$ and find a closed-form formula for ergodic rates. On the other hand, it is hard to find a closed-form formula for outage spectral efficiencies and Monte-Carlo method can be applied to evaluate it numerically.

\subsubsection{Ergodic Rate Calculation}

Using the SINR formula given in (\ref{SINR_eqn}), we can evaluate the ergodic rates as
\begin{equation}
\label{SE_ergodic_eqn1}
\text{SE}_{\text{ergodic},k,i}=\dfrac{1}{T_0}\mathbb{E}\left[\log_2\left(1+\dfrac{\displaystyle\sum_{n=1}^{N_k} \textbf{h}_{k,n,i}^H\textbf{h}_{k,n,i}P_t}{\displaystyle\sum_{j \not\in C_k} \displaystyle\sum_{n=1}^{N_k} \mathbb{E}\left[\textbf{h}_{k,n,j}^H\textbf{h}_{k,n,j}\right] P_t+\sigma_k^2}\right)\right].
\end{equation}
Notice that we use the factor $\dfrac{1}{T_0}$ since the symbol $s_i$ is transmitted within $T_0$ time/frequency samples. We can evaluate the expectation in the denominator part by defining \\
$b_{k,n,j}=\mathbb{E}[\textbf{h}_{k,n,j}^H\textbf{h}_{k,n,j}]$ and the resulting user ergodic spectral efficiency becomes
\begin{equation}
\label{SE_ergodic_eqn2}
\text{SE}_{\text{ergodic},k}=\displaystyle\sum_{i \in S_k} \text{SE}_{\text{ergodic},k,i} = \dfrac{1}{T_0}\displaystyle\sum_{i \in S_k}\mathbb{E}\left[\log_2\left(1+\dfrac{\displaystyle\sum_{n=1}^{N_k} \textbf{h}_{k,n,i}^H\textbf{h}_{k,n,i}P_t}{\displaystyle\sum_{j \not\in C_k} \displaystyle\sum_{n=1}^{N_k} b_{k,n,j} P_t+\sigma_k^2}\right)\right].
\end{equation}
To evaluate the spectral efficiency in (\ref{SE_ergodic_eqn2}), we need the pdf of $\displaystyle\sum_{n=1}^{N_k} \textbf{h}_{k,n,i}^H\textbf{h}_{k,n,i}$. To make the necessary calculations, we can use Theorem 1: 

\textbf{Theorem 1.}

Let $\textbf{x} \sim \mathcal{C}\mathcal{N}(\textbf{0}, \textbf{R}_x)$ be a random vector. $y=\textbf{x}^H\textbf{x}$ is a hypo-exponential random variable and its pdf can be written as
\begin{equation}
\label{SE_ergodic_eqn3}
\text{pdf}(y) = \left(\displaystyle\prod_{j=1}^a\lambda_j^{u_j}\right)\displaystyle\sum_{k=1}^a\displaystyle\sum_{\ell=1}^{u_k}\dfrac{\Phi_{k,\ell}(-\lambda_k)y^{u_k-\ell}e^{-\lambda_k y}}{(u_k-\ell)!(\ell-1)!}
\end{equation}
where the eigenvalues of $\textbf{R}_x^{-1}$ are 
\begin{equation}
\label{SE_ergodic_eqn4}
\underbrace{\lambda_1, \lambda_1, \ldots, \lambda_1}_{u_1}, \underbrace{\lambda_2, \lambda_2, \ldots, \lambda_2}_{u_2}, \ldots, \underbrace{\lambda_a, \lambda_a, \ldots, \lambda_a}_{u_a}
\end{equation}
and $\lambda_1, \lambda_2, \ldots, \lambda_a$ are pairwise distinct,
\begin{equation}
\label{SE_ergodic_eqn5}
\begin{aligned}
\Phi_{k,\ell}(-\lambda_k) &= (-1)^{\ell-1}(\ell-1)!\displaystyle\sum_{\Omega_{k, \ell}} \displaystyle\prod_{j=1, \: j \neq k}^a \dbinom{u_j+i_j-1}{u_j-1}(\lambda_j-\lambda_k)^{-u_j-i_j}, \\
\Omega_{k, \ell} &= \{\{i_j\}_{j=1, \: j \neq k}^a : \displaystyle\sum_{j=1, \: j \neq k}^a i_j =\ell-1, i_j \geq 0, \: \forall j \neq k\}.
\end{aligned}
\end{equation}

\textbf{Proof.}

When $\textbf{x}$ is white, i.e., $\textbf{R}_x$ is diagonal, the variable $y$ becomes a sum of independent exponential random variables with possibly different means. $y$ is called hypo-exponential random variable and for this case, the proof can be done using \cite{Ama97}. For the general case, we can again write $y$ as a sum of independent exponential random variables using a transformation
\begin{equation}
\label{SE_ergodic_eqn6}
\textbf{z}=\textbf{U}_x^H\textbf{R}_x^{-1/2}\textbf{x}
\end{equation}
where $\textbf{U}_x$ is a unitary matrix including the eigenvectors of $\textbf{R}_x=\textbf{U}_x\Lambda_x\textbf{U}_x^H$. The transformed variable $\textbf{z}$ is white as 
\begin{equation}
\label{SE_ergodic_eqn7}
\begin{aligned}
\mathbb{E}[\textbf{z}\textbf{z}^H]&=\mathbb{E}[\textbf{U}_x^H\textbf{R}_x^{-1/2}\textbf{x}\textbf{x}^H\textbf{R}_x^{-1/2}\textbf{U}_x] \\ &=\textbf{U}_x^H\textbf{R}_x^{-1/2}\mathbb{E}[\textbf{x}\textbf{x}^H]\textbf{R}_x^{-1/2}\textbf{U}_x = \textbf{U}_x^H\textbf{R}_x^{-1/2}\textbf{R}_x\textbf{R}_x^{-1/2}\textbf{U}_x=\textbf{I}.
\end{aligned}
\end{equation}
In this case, we can write $\textbf{x}^H\textbf{x}$ as
\begin{equation}
\label{SE_ergodic_eqn8}
\textbf{x}^H\textbf{x} = \textbf{z}^H\textbf{U}_x^H\textbf{R}_x\textbf{U}_x = \textbf{z}^H\Lambda_x \textbf{z} = \displaystyle\sum_{i=1}^{N_x} \mu_i|z_i|^2 
\end{equation}
where $\Lambda_x = \text{diag}(\mu_1, \mu_2, \ldots, \mu_{N_x})$. The variable $\mu_i|z_i|^2$ is exponential with mean $\mu_i$ and by the observation above, $z_i$'s are mutually independent. Therefore, the sum $\displaystyle\sum_{i=1}^{N_x} \mu_i|z_i|^2$ is a sum of independent exponentially distributed random variables with parameters $1/\mu_i$. In other words, $\lambda_i$'s are inverses of $\mu_i$'s which are eigenvalues of $\textbf{R}_x^{-1}$.

Using the covariance matrix $\textbf{h}_{\text{all}}$, we can find the sub-matrix related to the channels $\textbf{h}_{k,n,i}$ for $n=1, 2, \ldots, N_k$ and $i \in S_k$ and using Theorem 2, we can evaluate the pdf of nominator in (\ref{SE_ergodic_eqn2}). (Notice that the denominator in (\ref{SE_ergodic_eqn2}) is constant.)

To evaluate the expectation in (\ref{SE_ergodic_eqn2}), we will prove yet another theorem given by Theorem 2:

\textbf{Theorem 2.}

Let $X$ be a positive valued random variable with pdf $\text{pdf}(x)=\dfrac{x^{u-1}\lambda^ue^{-\lambda x}}{(u-1)!}, \: \forall x \geq 0$ where $u$ is a positive integer and $\lambda$ is a positive real number. Then we have
\begin{equation}
\label{SE_ergodic_eqn9}
\begin{aligned}
	\mathbb{E}\left[\log_2(1+X)\right] &= \int_0^{\infty} \log_2(1+x)\dfrac{x^{u-1}\lambda^ue^{-\lambda x}}{(u-1)!} \: dx \\
	&= \log_2(e)\left[(-1)^uP_u(\lambda)e^{\lambda}\text{Ei}(-\lambda)+Q_u(\lambda)\right]
	\end{aligned}
\end{equation}
where $P_i(x)$ and $Q_i(x)$ are two polynomials recursively defined by
\begin{equation}
\label{SE_ergodic_eqn10}
\begin{aligned}
P_{i+1}(x)&=\left(\dfrac{x}{i}-1\right)P_i(x)+\dfrac{x}{i}\dfrac{d}{dx}P_i(x), \quad i=1, 2, \ldots, \quad P_1(x)=1 \\
Q_{i+1}(x)&=Q_i(x)-(-1)^i\dfrac{1}{i}P_i(x)-\dfrac{x}{i}\dfrac{d}{dx}Q_i(x), \quad i=1, 2, \ldots, \quad Q_1(x)=0 
\end{aligned}
\end{equation}
and $\text{Ei}(\cdot)$ shows the exponential integral defined by
\begin{equation}
\label{SE_ergodic_eqn11}
\text{Ei}(x) = \int_{-\infty}^x \dfrac{e^t}{t} \: dt, \quad \forall x<0.
\end{equation}

\textbf{Proof.}

We prove the theorem using induction on $u$. For $u=1$ we have
\begin{equation}
\label{SE_ergodic_eqn12}
\begin{aligned}
\int_{0}^{\infty} \log_2(1+x)\lambda e^{-\lambda x}\: dx &\overset{\text{integration by parts}}{=} \left[\log_2(1+x)(-e^{-\lambda x})\right]_0^{\infty}-\int_0^{\infty}(-e^{-\lambda x})\dfrac{\log_2(e)}{1+x}\: dx \\
 &= \log_2(e)\int_{0}^{\infty}\dfrac{e^{-\lambda x}}{1+x} \: dx \overset{y=-\lambda(x+1)}{=} -\log_2(e)e^{\lambda} \int_{-\infty}^{-\lambda} \dfrac{e^y}{y} \: dy \\
&= -\log_2(e)e^{\lambda}\text{Ei}(-\lambda),
\end{aligned}
\end{equation}
as desired. Now assume that the formula is true for $i=u$ and we aim to prove that it is also true for $i=u+1$. Define 
\begin{equation}
\label{SE_ergodic_eqn13}
f_u(\lambda)=\int_0^{\infty} \log_2(1+x)\dfrac{x^{u-1}\lambda^ue^{-\lambda x}}{(u-1)!} \: dx.
\end{equation}
We obtain that 
\begin{equation}
\label{SE_ergodic_eqn14}
\begin{aligned}
\dfrac{d}{d\lambda}f_u(\lambda) &= \int_0^{\infty}\log_2(1+x)\dfrac{x^{u-1}}{(u-1)!}(u\lambda^{u-1}-x\lambda^u)e^{-\lambda x} \: dx \\
&= \dfrac{u}{\lambda}\left(f_u(\lambda)-f_{u+1}(\lambda)\right)
\end{aligned}
\end{equation}
and hence we get 
\begin{align}
\label{SE_ergodic_eqn15}
f_{u+1}(\lambda) &=f_u(\lambda) - \dfrac{\lambda}{u}\dfrac{d}{d\lambda}f_u(\lambda) \notag \\
&=\log_2(e)\left[(-1)^uP_u(\lambda)e^{\lambda}\text{Ei}(-\lambda)+Q_u(\lambda)\right] \notag \\
&\quad- \log_2(e)\dfrac{\lambda}{u}\dfrac{d}{d\lambda}\left[(-1)^uP_u(\lambda)e^{\lambda}\text{Ei}(-\lambda)+Q_u(\lambda)\right] \notag \\
&= \log_2(e)\left[(-1)^uP_u(\lambda)e^{\lambda}\text{Ei}(-\lambda)+Q_u(\lambda)\right] \notag \\
&\quad- \log_2(e)\dfrac{\lambda}{u}\left[(-1)^uP_u(\lambda)e^{\lambda}\text{Ei}(-\lambda)+(-1)^ue^{\lambda}\dfrac{d}{d\lambda}P_u(\lambda)\text{Ei}(-\lambda)\right] \\
&\quad - \log_2(e)\dfrac{\lambda}{u}\left[(-1)^ue^{\lambda}P_u(\lambda)\dfrac{e^{-\lambda}}{\lambda}+\dfrac{d}{d\lambda}Q_u(\lambda)\right] \notag \\
&= \log_2(e)(-1)^{u+1}e^{\lambda}\left[\left(\dfrac{\lambda}{u}-1\right)P_u(\lambda)+\dfrac{\lambda}{u}\dfrac{d}{d\lambda}P_u(\lambda)\right]\text{Ei}(-\lambda) \notag \\
&\quad + \log_2(e)\left[Q_u(\lambda)-(-1)^u\dfrac{1}{u}P_u(\lambda)-\dfrac{\lambda}{u}\dfrac{d}{d\lambda}Q_u(\lambda)\right] \notag \\
&= \log_2(e)\left[(-1)^{u+1}e^{\lambda}P_{u+1}(\lambda)\text{Ei}(-\lambda)+Q_{u+1}(\lambda)\right] \notag
\end{align}
and we are done.

Using Theorem 2 and 3, for $\textbf{x} \sim \mathcal{C}\mathcal{N}(\textbf{0}, \textbf{R}_x)$, we can calculate the mean $\mathbb{E}[\log_2(1+y)]$ for $y=\textbf{x}^H\textbf{x}$ as 
\begin{equation}
\label{SE_ergodic_eqn16}
\begin{aligned}
\mathbb{E}[\log_2(1+y)] &= \int_0^{\infty} \log_2(1+y) \left(\displaystyle\prod_{j=1}^a\lambda_j^{u_j}\right)\displaystyle\sum_{k=1}^a\displaystyle\sum_{\ell=1}^{u_k}\dfrac{\Phi_{k,\ell}(-\lambda_k)y^{u_k-\ell} e^{-\lambda_k y}}{(u_k-\ell)!(\ell-1)!} \: dy \\
&= \left(\displaystyle\prod_{j=1}^a\lambda_j^{u_j}\right)\displaystyle\sum_{k=1}^a\displaystyle\sum_{\ell=1}^{u_k} \int_0^{\infty} \log_2(1+y) \dfrac{\Phi_{k,\ell}(-\lambda_k)y^{u_k-\ell} e^{-\lambda_k y}}{(u_k-\ell)!(\ell-1)!} \: dy \\
&= \left(\displaystyle\prod_{j=1}^a\lambda_j^{u_j}\right)\displaystyle\sum_{k=1}^a\displaystyle\sum_{\ell=1}^{u_k} \dfrac{\Phi_{k,\ell}(-\lambda_k)\log_2(e)}{\lambda_k^{u_k-\ell+1}(\ell-1)!}\big[(-1)^{u_k-\ell+1}e^{\lambda_k}P_{u_k-\ell+1}(\lambda_k)\text{Ei}(-\lambda_k)\\
&+Q_{u_k-\ell+1}(\lambda_k)\big].
\end{aligned}
\end{equation}
To evaluate the ergodic rate of Alamouti-like orthogonal codes, we need to find $\lambda_k$'s and the corresponding values can be found using channel correlation matrix $\textbf{R}_{\text{all}}$. Firstly, notice that by the symmetry of the channel matrices $\textbf{h}_{k,n,i}$ we have 
\begin{equation}
\label{SE_ergodic_eqn17}
\textbf{h}_{k,n,i}^H\textbf{h}_{k,n,i}=\displaystyle\sum_{m \in U_k} |h_{m,k,n}|^2, \quad \forall i \in S_k.
\end{equation}
Therefore we need to consider the sub-matrix of $\textbf{R}_{\text{all}}$ corresponding to the channels $h_{m,k,n}$ for $m \in U_k, \: n=1, 2, \ldots, N_k$. Let the corresponding sub-matrix be $\textbf{R}_k$. The inverses of the eigenvalues of $\textbf{R}_k$ should be multiplied by the factor $\dfrac{1}{P_t}\left(\displaystyle\sum_{j \not\in C_k} \displaystyle\sum_{n=1}^{N_k} b_{k,n,j} P_t+\sigma_k^2\right)$ for each $i \in S_k$ to find corresponding $\lambda$ values. Notice that this factor can be evaluated using diagonal elements of $\textbf{R}_k$ and the order of $s_i$ in the orthogonal code used in cluster $C_k$.

Let the corresponding $\lambda$ values for $i$-th symbol of UE $k$ be 
\begin{equation}
\label{SE_ergodic_eqn18}
\underbrace{\lambda_{1,i}, \lambda_{1,i}, \ldots, \lambda_{1,i}}_{u_1}, \underbrace{\lambda_{2,i}, \lambda_{2,i}, \ldots, \lambda_{2,i}}_{u_2}, \ldots, \underbrace{\lambda_{a,i}, \lambda_{a,i}, \ldots, \lambda_{a,i}}_{u_a}
\end{equation}
where $\lambda_{1,i}, \lambda_{2,i}, \ldots, \lambda_{a,i}$ are pairwise distinct. Then we can evaluate the ergodic rate of UE $k$ as 
\begin{equation}
\label{SE_ergodic_eqn19}
\begin{aligned}
\text{SE}_{\text{ergodic},k}&=\dfrac{1}{T_0}\displaystyle\sum_{i \in S_k} \left(\displaystyle\prod_{r=1}^a\lambda_{r,i}^{u_r}\right)\displaystyle\sum_{r=1}^a\displaystyle\sum_{\ell=1}^{u_r} \dfrac{\Phi_{r,\ell,i}(-\lambda_{r,i})\log_2(e)}{\lambda_{r,i}^{u_r-\ell+1}(\ell-1)!}\big[(-1)^{u_r-\ell+1}e^{\lambda_{r,i}}P_{u_r-\ell+1}(\lambda_{r,i})\text{Ei}(-\lambda_{r,i})\\
&+Q_{u_r-\ell+1}(\lambda_{r,i})\big]
\end{aligned}
\end{equation}
where
\begin{equation}
\label{SE_ergodic_eqn20}
\begin{aligned}
\Phi_{r,\ell,i}(-\lambda_{r,i}) &= (-1)^{\ell-1}(\ell-1)!\displaystyle\sum_{\Omega_{r, \ell}} \displaystyle\prod_{j=1, \: j \neq r}^a \dbinom{u_j+i_j-1}{u_j-1}(\lambda_{j,i}-\lambda_{r,i})^{-u_j-i_j}, \\
\Omega_{r, \ell} &= \{\{i_j\}_{j=1, \: j \neq r}^a : \displaystyle\sum_{j=1, \: j \neq r}^a i_j =\ell-1, i_j \geq 0, \: \forall j \neq r\}, \\
P_{j+1}(x)&=\left(\dfrac{x}{j}-1\right)P_j(x)+\dfrac{x}{j}\dfrac{d}{dx}P_j(x), \quad j=1, 2, \ldots, \quad P_1(x)=1 \\
Q_{j+1}(x)&=Q_j(x)-(-1)^j\dfrac{1}{j}P_j(x)-\dfrac{x}{j}\dfrac{d}{dx}Q_j(x), \quad j=1, 2, \ldots, \quad Q_1(x)=0 \\
\text{Ei}(x) &= \int_{-\infty}^x \dfrac{e^t}{t} \: dt, \quad \forall x<0.
\end{aligned}
\end{equation}
The polynomials $P_j(x)$ and $Q_j(x)$ can be evaluated offline for values $j=1, 2, \ldots, j_0$ where $j_0$ denotes the maximum possible dimension of $\textbf{R}_k$ which is equal to $4 \max_k N_k$. Notice that according to our code period limit 8, each user can be served by at most 4 RU antennas in total. The function $\Phi_{k,\ell,i}$ involves multiplications of pairwise differences of $\lambda_{r,i}$'s and other constant coefficients (including binomial ones) can also be calculated offline. The only non-elementary function in (\ref{SE_ergodic_eqn20}) is the $\text{Ei}$ function and it can be evaluated by look-up table.

\subsubsection{Outage Rate Calculation}

To evaluate outage rates (or spectral efficiencies), we use the SINR equations obtained in (\ref{SINR_eqn}). It is hard to find a closed form formula as it is required to find the inverse image of the CDF of SINRs. To find a solution, we use Monte-Carlo method by generating independent channel coefficients and approximate the probability given in (\ref{SE_outage_eqn2}). The steps of the evaluation process is as follows:

\textbf{Step 1:} Using the correlation matrices $\textbf{R}_k$ for all $k$, generate $N_{\text{trial}}$ different channel coefficients $\textbf{h}_{k,n,i}$ for all $k, n, i$ triples.

\textbf{Step 2:} For each channel realization, calculate the nominator of SINRs using (\ref{SINR_eqn}). Using all channel realizations calculate the mean terms in the denominator of (\ref{SINR_eqn}). This step gives us $N_{\text{trial}}$ different SINR values for each $(k, i)$ pair.

\textbf{Step 3:} Using SINR realizations found after Step 2, calculate the CDF of SINRs for each $(k, i)$ pair and find the SINR threshold value for which $\text{Pr}(\text{SINR}_{k,i}<\text{SINR}_{\text{min}, k, i})=P_{\text{out}}$ where $P_{\text{out}}$ is a constant outage probability value chosen before.

\textbf{Step 4:} Evaluate the outage spectral efficiency as 
\begin{equation}
\text{SE}_{\text{outage},k}=(1-P_{\text{out}})\displaystyle\sum_{i \in S_k} \text{SE}_{\text{outage},k, i} = (1-P_{\text{out}})\displaystyle\sum_{i \in S_k} \log_2(1+\text{SINR}_{\text{min}, k, i}), \quad \forall k.
\end{equation}
The outage spectral efficiencies can be approximately evaluated using Step 1-4, and a higher $N_{\text{trial}}$ value results in a better accuracy.

\section{Clustering for Orthogonal Coding in D-MIMO}

The main task in this study is to determine RU-UE clusters to optimize overall system performance. In this section, we present a heuristic clustering method relying on closed-form ergodic rates where clusters are formed in 3 stages. The algorithm uses large-scale fading and correlation parameters and can be implemented at CP. We also investigate the asymptotic complexity of the proposed method to observe its practical implementation cost.

\subsection{Cluster Formation}

The steps of the algorithm are given below.

\begin{itemize}
\item \textbf{Step 1 (One-to-one matching):} Sort users according to the $\max_m \beta_{m,k}$. The most prior one has the minimal $\max_m \beta_{m,k}$. Then starting from the most prior one match UEs and RUs in one-to-one manner where each UE is matched with the RU with the largest possible fading coefficient. This operation forms $K$ disjoint clusters $C_1, C_2, \ldots, C_K$. At this step, each user is associated with one antenna of different RUs.
\item \textbf{Step 2 (Cluster merging):} Prioritize clusters according to the minimum spectral efficiency $\text{SE}_{\text{min}}$ in each cluster. The most prior one has the minimal $\text{SE}_{\text{min}}$. Starting from the most prior cluster pair, check whether merging the two clusters increases the worst $K/4$ spectral efficiencies of users. While performing each check, we re-evaluate each ergodic rate. After a merging occurs, we perform prioritization of remaining clusters again and continue the same process. Each merging decreases the number of total clusters and hence after finitely many trials we obtain a stable clustering.
\item \textbf{Step 3 (Add remaining antennas):} In this step, we try to add unused RU antennas to clusters obtained after Step 2. According to the final prioritization of clusters obtained in Step 2, we check whether any of unused RU antennas can be added to the most prior cluster to increase the worst $K/4$ spectral efficiencies of users. If any RU antenna is added to a cluster, then we re-calculate the prioritization of clusters and continue this process until all unused RU antennas are checked. 
\end{itemize}

Notice that Step 1-2 forms single-user clusters only. Multi-user clusters are obtained after Step 3. In Fig. 6, we present an example output of the clustering algorithm. We observe that the algorithm forms three clusters with various types. It is shown that isolated users are served by single-user clusters whereas closely separated users are served in the same cluster. 

\begin{figure}[ht]
\centering
  \includegraphics[width=0.8\linewidth]{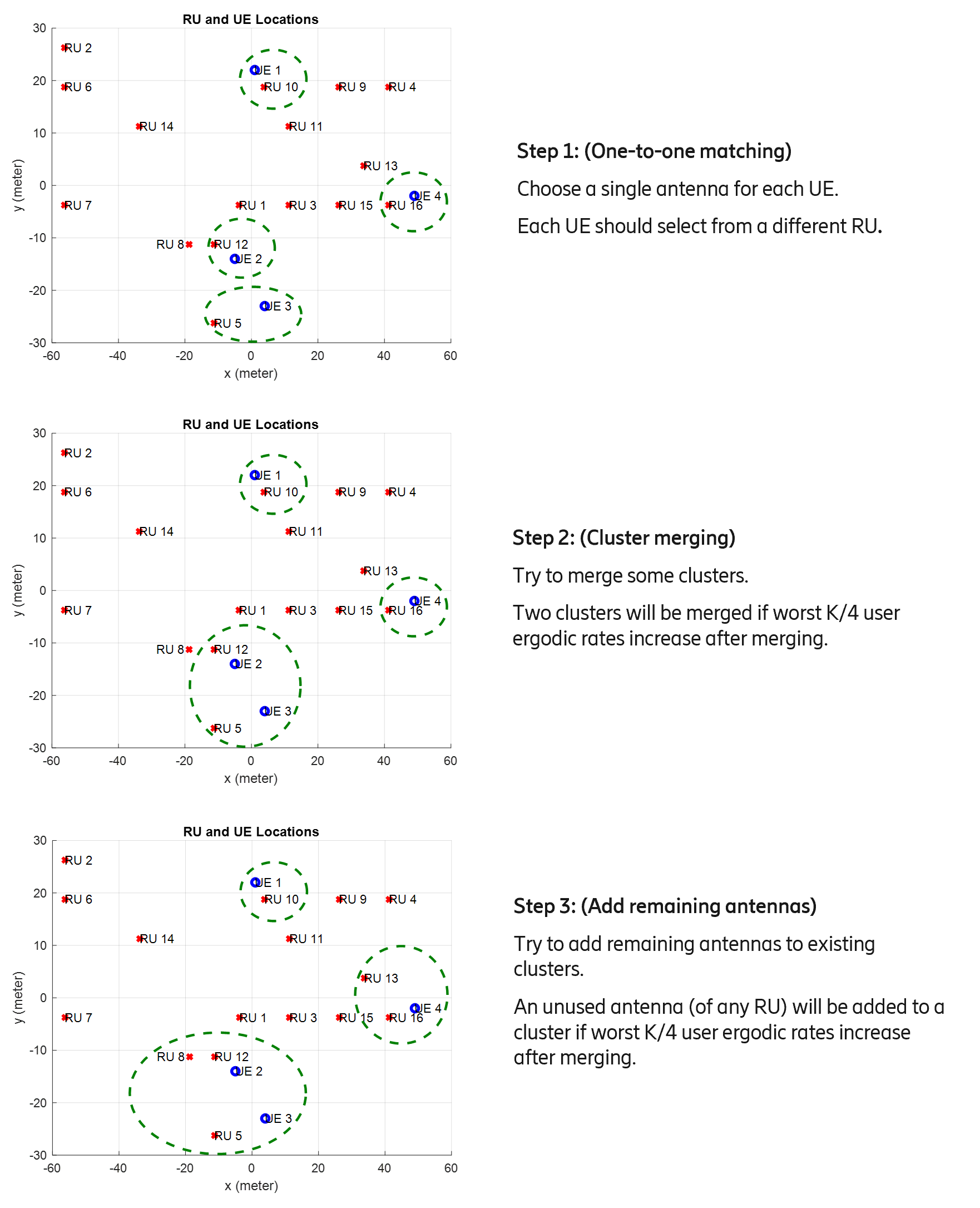}
  \caption{An example clustering scheme}
  \label{fig6}
\end{figure}

\subsection{Asymptotic Complexity of the Clustering}

We can analyze the complexity of each step separately. We assume that $L_m=L$ for all $m$.

\textbf{Step 1:} This step involves initial selection, and its complexity is negligible. 

\textbf{Step 2:} The maximum number of cluster merging trials is equal to
\begin{equation}
\label{complexity_eqn1}
\dbinom{K}{2}+\dbinom{K-1}{2}+\cdots+\dbinom{C}{2} = \dbinom{K+1}{3} - \dbinom{C}{3}
\end{equation}
where $C$ is the final value of the number of clusters at the end of Step 2.  

\textbf{Step 3:} The maximum number of antenna addition trials is given by
\begin{equation}
\label{complexity_eqn2}
(ML-K)C.
\end{equation}
The complexity of each trial in Step 2 and 3 is $\mathcal{O}(K)$ as we evaluate the new values of all user rates. So, the overall complexity becomes
\begin{equation}
\label{complexity_eqn3}
\mathcal{O}\left(\left(\dbinom{K+1}{3} - \dbinom{C}{3} + (ML-K)C\right)K\right).
\end{equation}
Consider the function $f(C)=\left(\dbinom{K+1}{3} - \dbinom{C}{3} + (ML-K)C\right)$. The first and second derivatives of $f$ are given by 
\begin{equation}
\label{complexity_eqn4}
f'(C)=ML-K-\dfrac{1}{6}(3C^2-6C+2), \quad f''(C)=1-C
\end{equation}
and hence $f$ is concave and assuming $ML>>K$, and using $K \geq C$, we get $ML-K>\dfrac{1}{6}(3C^2-6C+2)$ and hence $f$ takes its maximum value when $C=K$. In this case, we get 
\begin{equation}
\label{complexity_eqn5}
f(K)=MLK-\dfrac{K^2+K}{2}.
\end{equation}
As $ML>>K$, the term $\dfrac{K^2+K}{2}$ can be ignored and the asymptotic complexity becomes $\mathcal{O}(MLK^2)$. 

\section{Other Methods for Comparison}

To make a comparison, we consider different baseline methods. Before explaining each method separately, we will first write general received signal, SINR and achievable outage and ergodic spectral efficiency equations. In general, the signal transmitted from RU $m$ can be written as

\begin{equation}
\label{other_methods_eqn1}
\textbf{x}_m = \displaystyle\sum_{k=1}^K \sqrt{P_t\eta_{m,k}}\textbf{W}_{m,k} \textbf{q}_k
\end{equation}

where $\textbf{q}_k \in \mathbb{C}^{N_k \times 1}$ is the symbol vector transmitted to UE $k$, $P_t$ is the maximum transmit power of each RU antenna, $\textbf{W}_{m,k} \in \mathbb{C}^{L_m \times N_k}$ is the precoding matrix and $\eta_{m,k}$ is the power control coefficient for the pair RU $m$ and UE $k$. According to the channel information at RUs and the number of antennas of the UEs, we can transmit a single layer or more than one layer to the users.

\noindent \textbf{Case 1.1: Single-Layer Transmission}

This type of transmission is used for methods without CSI at Tx side, or when a UE has a single antenna. In this case, the vector $\textbf{q}_k$ involves the same symbol $s_k$, i.e., $\textbf{q}_k=[s_k \: s_k \: \cdots \: s_k]^T$ where $s_k$ has zero mean and unity variance, and we do not apply any precoding. The transmitted signal by the $m$-th RU becomes
\begin{equation}
\label{other_methods_eqn2}
\displaystyle\sum_{k=1}^K \sqrt{P_t\eta_{m,k}} \begin{bmatrix} s_k \\ s_k \\ \vdots \\ s_k \end{bmatrix} = \displaystyle\sum_{k=1}^K \sqrt{P_t\eta_{m,k}} \textbf{1}_{L_m} s_k.
\end{equation}

\noindent \textbf{Case 1.2: Multi-Layer Transmission}

This type of transmission is used for methods with Tx CSI and UEs with multiple antennas. In this case, the vector $\textbf{q}_k$ involves $N_k$ independent, identically distributed entries with mean zero and variance $1$. The transmitted signal by the $m$-th RU can be written as in (\ref{other_methods_eqn1}).

In general, the received signal by the $k$-th UE can be expressed as
\begin{equation}
\label{other_methods_eqn3}
\textbf{r}_k = \displaystyle\sum_{m=1}^M \textbf{H}_{m,k}\textbf{x}_m + \textbf{z}_k
\end{equation}
where $\textbf{H}_{m, k} \in \mathbb{C}^{N_k \times L_m}$ is the channel matrix between RU $m$ and UE $k$, $\textbf{z}_k$ is the receiver noise vector for UE $k$ whose entries are independent, circularly symmetric, Gaussian with standard deviation $\sigma_k^2$. 

Using the structure of $\textbf{q}_k$, the received signal by the $k$-th UE can be expressed as
\begin{equation}
\label{other_methods_eqn4}
\textbf{r}_k = \begin{cases} \displaystyle\sum_{\ell=1}^K\displaystyle\sum_{m=1}^M\sqrt{P_t\eta_{m, \ell}} \textbf{H}_{m, k}\textbf{1}_{L_m}s_{\ell} + \textbf{z}_k, \quad \text{single-layer} \\
\displaystyle\sum_{\ell=1}^K\displaystyle\sum_{m=1}^M\sqrt{P_t\eta_{m, \ell}} \textbf{H}_{m, k}\textbf{W}_{m, \ell}\textbf{q}_{\ell} + \textbf{z}_k, \quad \text{multi-layer} 
\end{cases} 
\end{equation}
We can rewrite $\textbf{r}_k$ as
\begin{equation}
\label{other_methods_eqn5}
\textbf{r}_k = \begin{cases} \sqrt{P_t}\displaystyle\sum_{\ell=1}^K \textbf{D}_{k,\ell,\text{sl}}s_{\ell} + \textbf{z}_k, \quad \text{single-layer}  \\
\sqrt{P_t}\displaystyle\sum_{\ell=1}^K \textbf{D}_{k,\ell,\text{ml}}\textbf{q}_{\ell} + \textbf{z}_k, \quad \text{multi-layer} 
\end{cases} 
\end{equation}
where $\textbf{D}_{k,\ell,\text{sl}} \in \mathbb{C}^{N_k \times 1}$ and $\textbf{D}_{k,\ell,\text{ml}} \in \mathbb{C}^{N_k \times N_{\ell}}$ are defined as
\begin{equation}
\label{other_methods_eqn6}
\begin{aligned}
\textbf{D}_{k,\ell,\text{sl}} &= \sqrt{P_t}\displaystyle\sum_{m=1}^M \sqrt{\eta_{m,\ell}}\textbf{H}_{m,k}\textbf{1}_{L_m}, \\
\textbf{D}_{k,\ell,\text{ml}} &= \sqrt{P_t}\displaystyle\sum_{m=1}^M \sqrt{\eta_{m,\ell}}\textbf{H}_{m,k}\textbf{W}_{m,\ell}. 
\end{aligned}
\end{equation}
We assume that there exists a separate power amplifier for each RU antenna and hence we consider per-antenna power transmit constraints that can be expressed as
\begin{equation}
\label{other_methods_eqn7}
\mathbb{E}[|[\textbf{x}_m]_i|^2] = [\mathbb{E}[\textbf{x}_m\textbf{x}_m^H]]_{i,i} \leq P_t, \quad 1 \leq m \leq  M, \: 1 \leq i \leq L_m
\end{equation}
where $[\textbf{x}]_i$ and $[\textbf{X}]_{i,i}$ denote the $i$-th element of the vector $\textbf{x}$ and $(i,i)$-th element of the diagonal matrix $\textbf{X}$, respectively. For single and multi layer transmissions, we can express the condition given in (\ref{other_methods_eqn7}) as
\begin{equation}
\label{other_methods_eqn8}
\begin{aligned}
\displaystyle\sum_{k=1}^K \eta_{m,k} \leq 1, &\quad 1 \leq m \leq M, \quad \text{single-layer} \\
\displaystyle\sum_{k=1}^K \eta_{m,k}[\mathbb{E}[\textbf{W}_{m,k}\textbf{W}_{m,k}^H]]_{i,i} \leq 1, &\quad 1 \leq m \leq M, \: 1 \leq i \leq L_m, \quad \text{multi-layer} 
\end{aligned}
\end{equation}
We also consider two different cases according to the channel state information at the receiver side.

\noindent \textbf{Case 2.1: Perfect CSI at Rx Side} 

In this case, we assume that each UE perfectly knows its effective channel. We can write the desired signal and the other signal parts as
\begin{equation}
\label{other_methods_eqn9}
\textbf{r}_k = \begin{cases} \underbrace{\sqrt{P_t}\textbf{D}_{k,k,\text{sl}} s_k}_{\text{desired}} + \underbrace{\sqrt{P_t}\displaystyle\sum_{\ell \neq k}^K \textbf{D}_{k,\ell,\text{sl}} s_{\ell}}_{\text{interference}} + \underbrace{\textbf{z}_k}_{\text{noise}},
\quad \text{single-layer} \\
\underbrace{\sqrt{P_t}\textbf{D}_{k,k,\text{ml}} \textbf{q}_k}_{\text{desired}} + \underbrace{\sqrt{P_t}\displaystyle\sum_{\ell \neq k}^K \textbf{D}_{k,\ell,\text{ml}} \textbf{q}_{\ell}}_{\text{interference}} + \underbrace{\textbf{z}_k}_{\text{noise}}, \quad \text{multi-layer} 
\end{cases}
\end{equation}
Using the information theoretic approach given in \cite{Mai20}, the achievable spectral efficiency for the $k$-th user can be written as
\begin{equation}
\label{other_methods_eqn10}
\text{SE}_{\text{ergodic}, k} = \begin{cases} 
\mathbb{E}\left[\log_2\left(1+\dfrac{P_t\textbf{D}_{k,k,\text{sl}}^H\textbf{D}_{k,k,\text{sl}}}{P_t\displaystyle\sum_{\ell\neq k}^K\mathbb{E}[\textbf{D}_{k,\ell,\text{sl}}^H\textbf{D}_{k,\ell,\text{sl}}]+\sigma_k^2}\right)\right], \quad \text{single-layer} \\
\mathbb{E}[\log_2(|\textbf{I}_{N_k}+P_t\textbf{D}_{k,k}^H\Psi_{k,k,1}^{-1}\textbf{D}_{k,k}|)], \quad \text{multi-layer}
\end{cases} 
\end{equation}
where $\Psi_{k,k,1}=P_t\displaystyle\sum_{\ell\neq k}^K\mathbb{E}[\textbf{D}_{k,\ell,\text{ml}}\textbf{D}_{k,\ell,\text{ml}}^H]+\sigma_k^2\textbf{I}_{N_k}$. 

\noindent \textbf{Case 2.2: Statistical CSI at Rx Side} 

In this case, we assume that each UE only knows the mean of its effective channel. Therefore, we can write the desired signal and the other signal parts as
\begin{equation}
\label{other_methods_eqn11}
\textbf{r}_k = \begin{cases} \underbrace{\sqrt{P_t}\overline{\textbf{D}}_{k,k,\text{sl}}s_k}_{\text{desired}} + \underbrace{\sqrt{P_t}(\textbf{D}_{k,k,\text{sl}}-\overline{\textbf{D}}_{k,k,\text{sl}})s_k}_{\text{mismatch}} + \underbrace{\sqrt{P_t}\displaystyle\sum_{\ell \neq k}^K \textbf{D}_{k,\ell,\text{sl}}s_{\ell}}_{\text{interference}} + \underbrace{\textbf{z}_k}_{\text{noise}}, \quad \text{single-layer} \\
\underbrace{\sqrt{P_t}\overline{\textbf{D}}_{k,k,\text{ml}}\textbf{q}_k}_{\text{desired}} + \underbrace{\sqrt{P_t}(\textbf{D}_{k,k,\text{ml}}-\overline{\textbf{D}}_{k,k,\text{ml}})\textbf{q}_k}_{\text{mismatch}} + \underbrace{\sqrt{P_t}\displaystyle\sum_{\ell \neq k}^K \textbf{D}_{k,\ell,\text{ml}}\textbf{q}_{\ell}}_{\text{interference}} + \underbrace{\textbf{z}_k}_{\text{noise}}, \quad \text{multi-layer} 
\end{cases} 
\end{equation}
Here $\overline{\textbf{D}}_{k,k,\text{sl}}=\mathbb{E}[\textbf{D}_{k,k,,\text{sl}}]$ and $\overline{\textbf{D}}_{k,k,\text{ml}}=\mathbb{E}[\textbf{D}_{k,k,,\text{ml}}]$ are the means of the effective channels, and the mismatch part includes the channel uncertainty due to the limited knowledge about the effective channel. Using the information theoretic approach given in \cite{Mai20} again, the achievable spectral efficiency for the $k$-th user can be written as
\begin{equation}
\label{other_methods_eqn12}
\text{SE}_{\text{ergodic}, k} = \begin{cases} \log_2\left(1+\dfrac{P_t\overline{\textbf{D}}_{k,k,\text{sl}}^H\overline{\textbf{D}}_{k,k,\text{sl}}}{P_t\displaystyle\sum_{\ell \neq k}^K \mathbb{E}[\textbf{D}_{k,\ell,\text{sl}}^H\textbf{D}_{k,\ell,\text{sl}}]+\sigma_k^2}\right), \quad \text{single-layer} \\
 \log_2(|\textbf{I}_{N_k}+P_t\overline{\textbf{D}}_{k,k,\text{ml}}^H\Psi_{k,k,2}^{-1}\overline{\textbf{D}}_{k,k,\text{ml}}|), \quad \text{multi-layer}
\end{cases} 
\end{equation}
where $\Psi_{k,k,2}=P_t\displaystyle\sum_{\ell \neq k}^K \mathbb{E}[\textbf{D}_{k,\ell,\text{ml}}\textbf{D}_{k,\ell,\text{ml}}^H]-P_t\overline{\textbf{D}}_{k,k,\text{ml}}\overline{\textbf{D}}_{k,k,\text{ml}}^H+\sigma_k^2\textbf{I}_{N_k}$.

In this study, we consider various baseline methods and we divide them into two categories according to the channel knowledge at the RU side.

\subsection{Methods with unknown instantaneous CSI at RU side}

We consider small-cells and single frequency networks (SFN) in this category.

\noindent \textbf{Small-cells:} Each UE is served by a single RU with the largest $\beta_{m,k}$ value. The corresponding UE symbol is transmitted from all antennas of the selected RU. We consider single layer transmission and assume perfect CSI at UE side. Let $m_k$ be the selected RU for the $k$-th user. We select the power control coefficients as 
\begin{equation}
\label{other_methods_eqn13}
\eta_{m,k} = \begin{cases} 1, \quad m=m_k \\
0, \quad m \neq m_k.
\end{cases}
\end{equation}

\noindent \textbf{SFN:} Each UE selects a set of RUs using $95\%$ rule which is defined as follows:

\noindent \underline{$95\%$ RU Selection Rule:} For a fixed user $k$, sort $\beta_{m,k}$'s in the descending order. Choose the first $M_k$ of them so that their sum is at least $95\%$ of the total sum. Select $M_k$ as small as possible.

The corresponding UE symbol is transmitted from all antennas of all selected RUs. In this method we again consider single layer transmission and assume perfect CSI at UE side. Let $m_1, m_2, \ldots, m_{M_k}$ be the selected RUs for the $k$-th user. We select the power control coefficients as 
\begin{equation}
\label{other_methods_eqn14}
\eta_{m,k} = \begin{cases} \dfrac{1}{|V_m|}, \quad m \in U_k \\
0, \quad \quad m \notin U_k
\end{cases}
\end{equation} 
where $U_k$ is the set of RUs serving UE $k$ and $V_m$ is the set of users served by RU $m$. Notice that this selection gives equal power to all associated users for each RU and satisfies the transmit power constraint in (\ref{other_methods_eqn8}).
 
\subsection{Methods with known instantaneous CSI at RU side}

We consider different variants of maximal ratio transmission (MRT) in this category.

\noindent \textbf{MRT ($95\%$):} Each UE selects a set of RUs using $95\%$ rule. The selected RUs apply MRT precoding to send the corresponding user symbols. We consider multi-layer transmission for multi-antenna users. The precoding matrix is chosen as 
\begin{equation}
\label{other_methods_eqn15}
\textbf{W}_{m,k} = \textbf{H}_{m,k}^H, \quad \forall m, k
\end{equation} 
and the power control coefficients are selected as 
\begin{equation}
\label{other_methods_eqn16}
\eta_{m,k} = \begin{cases} \dfrac{1}{\displaystyle\sum_{\ell \in V_m} N_{\ell}\beta_{m,\ell}}, \quad m \in U_k \\
0, \quad \quad m \notin U_k. 
\end{cases}
\end{equation} 
This selection satisfies the power constraint given in (\ref{other_methods_eqn8}) as 
\begin{equation}
\label{other_methods_eqn17}
[\mathbb{E}[\textbf{W}_{m,k}\textbf{W}_{m,k}^H]]_{i,i} = [\mathbb{E}[\textbf{H}_{m,k}^H\textbf{H}_{m,k}]]_{i,i} = N_k \beta_{m,k}.
\end{equation} 

\noindent \textbf{MRT (1 RU)}: Each UE is served by a single RU with the largest $\beta_{m,k}$ value. The selected RU applies MRT precoding to send the corresponding user symbols. We consider multi-layer transmission for multi-antenna users. The equations given in (\ref{other_methods_eqn15}) and  (\ref{other_methods_eqn16}) are valid with $|U_k|=1$ as each user is served by a single RU.

We consider both perfect and statistical CSI cases at Rx side for MRT ($95\%$) and MRT (1 RU) methods. Achievable ergodic spectral efficiency values can be calculated using (\ref{other_methods_eqn10}) and (\ref{other_methods_eqn12}).

\section{Numerical Results}

To see the performance benefits of the orthogonal codes, we perform various numerical simulations. We consider a smart factory area to deploy RUs and UEs. We define grid points for RUs and UEs and select the RU and UE positions randomly from the defined grid points. The simulation parameters are given in Table IV. 

\begin{table}[ht]
\caption{Simulation parameters}
\centering 
\begin{tabular}{| l | c |}
\hline
\textbf{Parameter} & \textbf{Value}/\textbf{Model} \\
\hline
Carrier frequency and bandwidth & $28$ GHz, $200$ MHz \\
\hline
Area & $120 \times 60$ meters \\
\hline
$P_t$ & $0.2$ W \\
\hline
RU grid & $16 \times 8$ grid with spacing $7.5$ meters \\
\hline
UE grid & $120 \times 60$ grid with spacing $1$ meter \\
\hline
Channel model & 3GPP InF-SL \cite{3gpp_InF_SL} \\
\hline
UE noise figure & $9$ dB \\
\hline
$P_{\text{out}}$ & $0.01$ \\
\hline
\end{tabular}
\label{table_4}
\end{table}

We assume that each RU has $L$ antennas and each UE has $N$ antennas. We consider the effects of the number of RUs ($M$), the number of UEs ($K$), the number of RU antennas ($L$), and the number of UE antennas ($N$) in the simulations. We compare outage spectral efficiencies of methods without Tx CSI, and ergodic spectral efficiencies of all methods.

\newpage

\begin{figure}[ht]	
	\centering
	\begin{subfigure}{0.5\linewidth}
		\centering
		\includegraphics[width=0.9\linewidth]{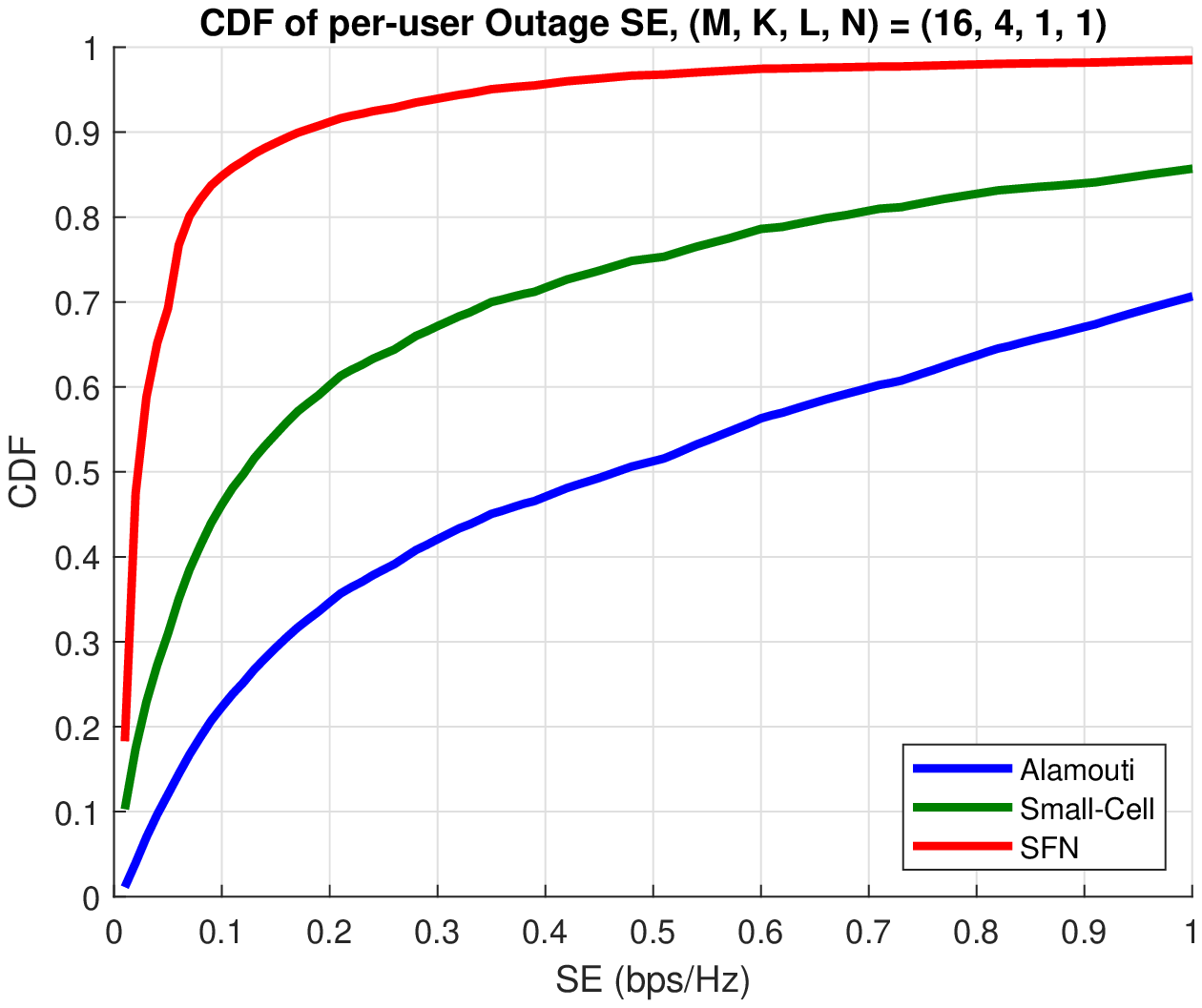}
		\caption{CDF of per-user Outage SE}
	\end{subfigure}%
	\begin{subfigure}{0.5\linewidth}
		\centering
		\includegraphics[width=0.9\linewidth]{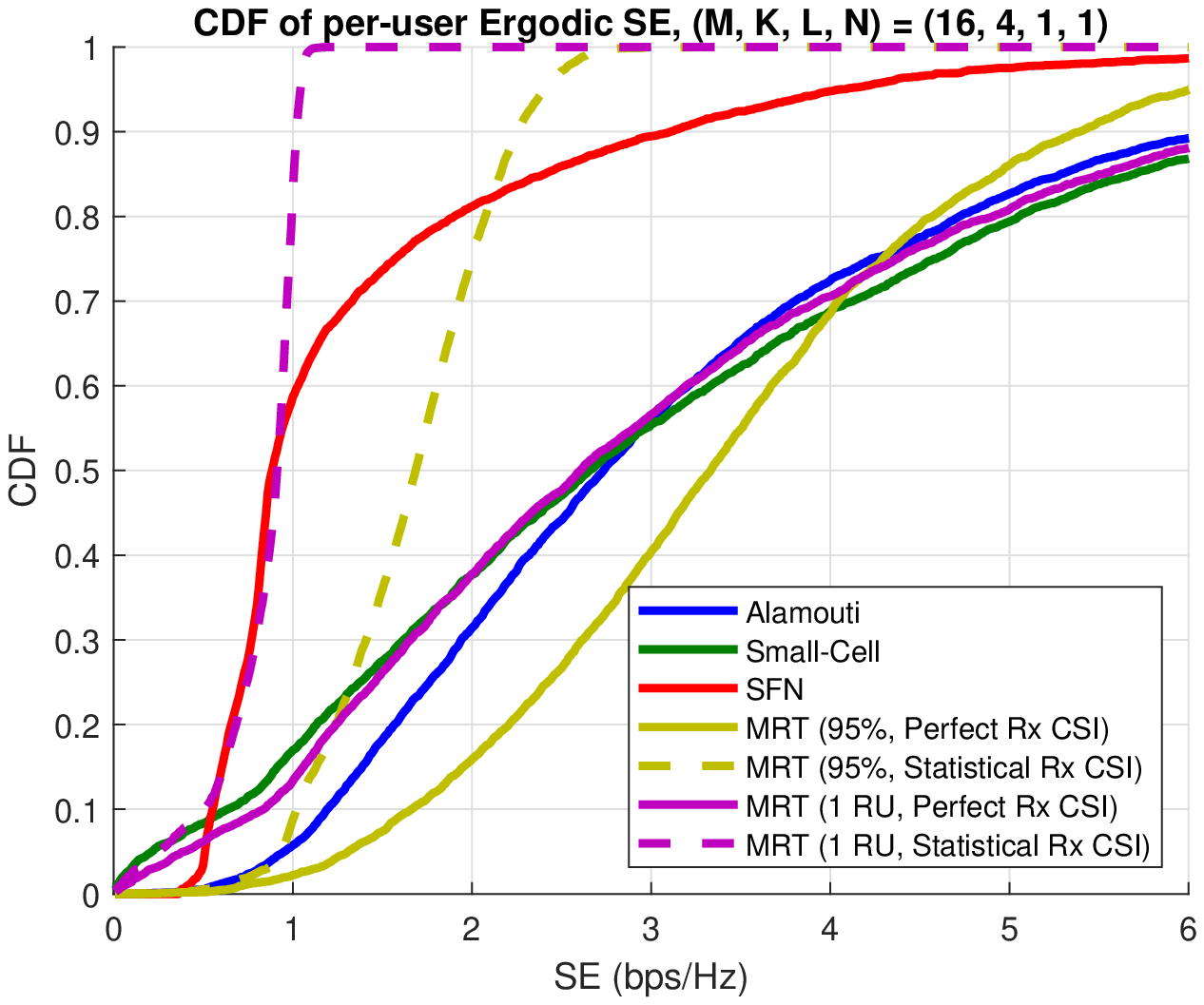}
		\caption{CDF of per-user Ergodic SE}
	\end{subfigure}
\caption{CDF of per-user Outage and Ergodic SEs ($M=16, K=4, N=L=1$).}
\label{fig7}	
\end{figure}

In Fig. 7, we see the per-user CDFs for both outage and ergodic SEs. We observe that orthogonal codes significantly outperforms small-cell and SFN in terms of outage rates. It has much better $5$-th percentile ergodic SEs compared to small-cell, SFN and MRT (1 RU), and better than SFN and MRT with statistical RX CSI in terms of median ergodic SEs. We conclude that for single antenna case, by the help of clustering, orthogonal coding can achieve even better performance than some methods with Tx CSI. Another observation is about comparison of ergodic SEs of orthogonal coding and small-cells. Notice that in Step 1 of the clustering algorithm, we match RUs and UEs as in small-cell approach. Step 2 and 3 tries to maximizes the worst case user ergodic SEs and hence we observe a significant enhancement on $5$-th percentile user ergodic SEs. On the other hand, the median values are similar for orthogonal coding and small-cell, and this shows that the clustering can optimize the worst case users by maintaining a similar performance for all users in average. The outage SEs for orthogonal coding is always better as adding more antennas to serve a user increases the reliability and decreases the effect of outage.

\newpage

\begin{figure}[ht]	
	\centering
	\includegraphics[width=\linewidth]{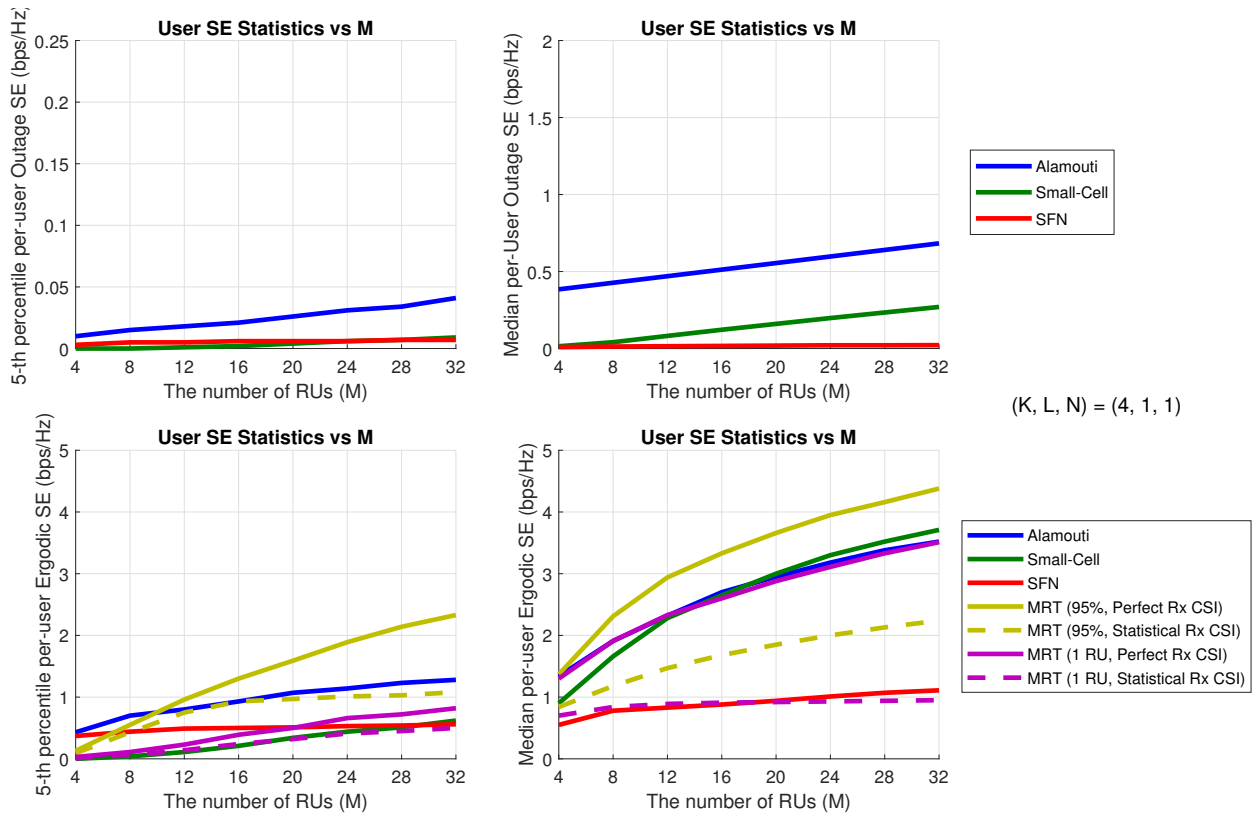}
\caption{$5$-th percentile and median SEs for various $M$ values ($K=4, L=N=1$).}
\label{fig8}	
\end{figure}

In Fig. 8, we present the $5$-th percentile and median SEs for different $M$ values to see the effects of the number of RUs. The performance in terms of both outage and ergodic SEs increases as the number of RUs increases. Orthogonal coding outperforms small-cell and SFN for all $M$ values in terms of outage rates. $5$-th percentile ergodic SEs of orthogonal coding are much better than those of small-cell and MRT (1 RU) whereas median ergodic SEs of these three methods are similar. In small-cell and MRT (1 RU) methods, the users are served by a single RU and hence increasing $M$ does not enhance the performance much. On the contrary, MRT ($95\%$) outperforms orthogonal coding for large $M$ values as more RUs can cooperate to serve users.

\newpage

\begin{figure}[ht]	
	\centering
	\includegraphics[width=0.9\linewidth]{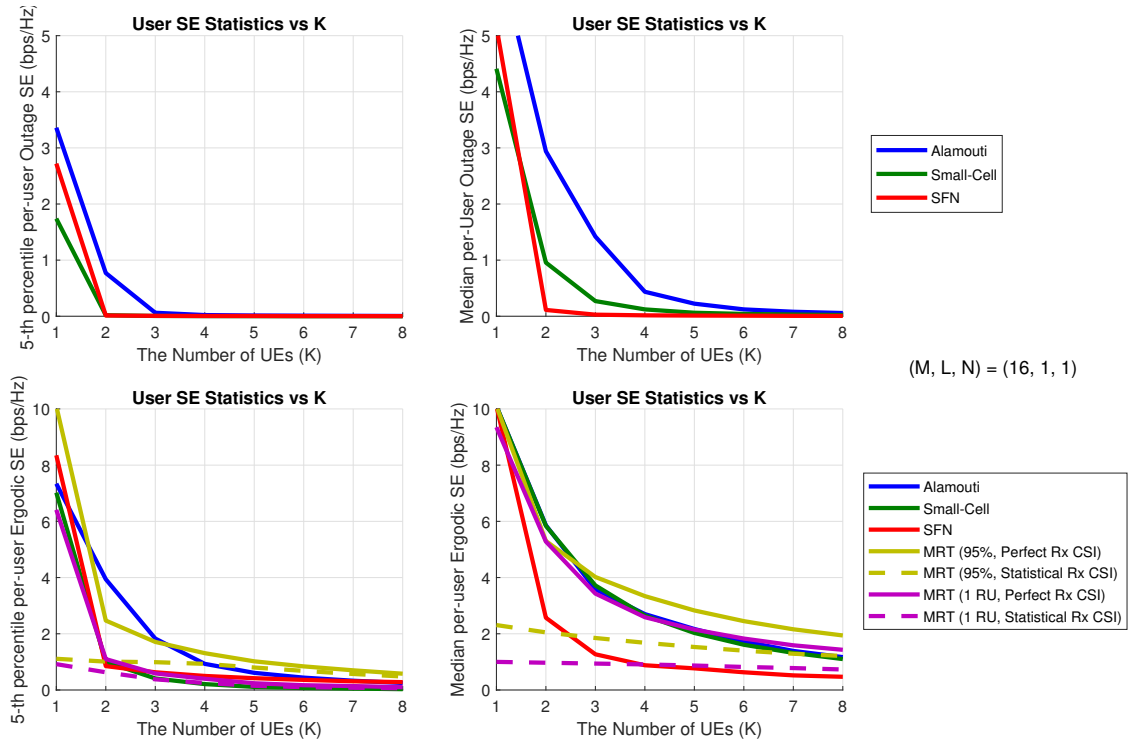}
\caption{$5$-th percentile and median SEs for various $K$ values ($M=16, L=N=1$).}
\label{fig9}	
\end{figure}

Fig. 9 shows the effect of the number of users. As in previous comparisons, orthogonal coding outperforms small-cell and SFN for all $K$ values in terms of outage SEs. For $K>4$, we observe a dramatic decrease in outage SEs. According to this result, we can conclude that Cluster Types 11 and 12 with 5 and 6 users, respectively do not provide satisfactory rate values. 

\newpage

\begin{figure}[ht]
	\centering
	\includegraphics[width=\linewidth]{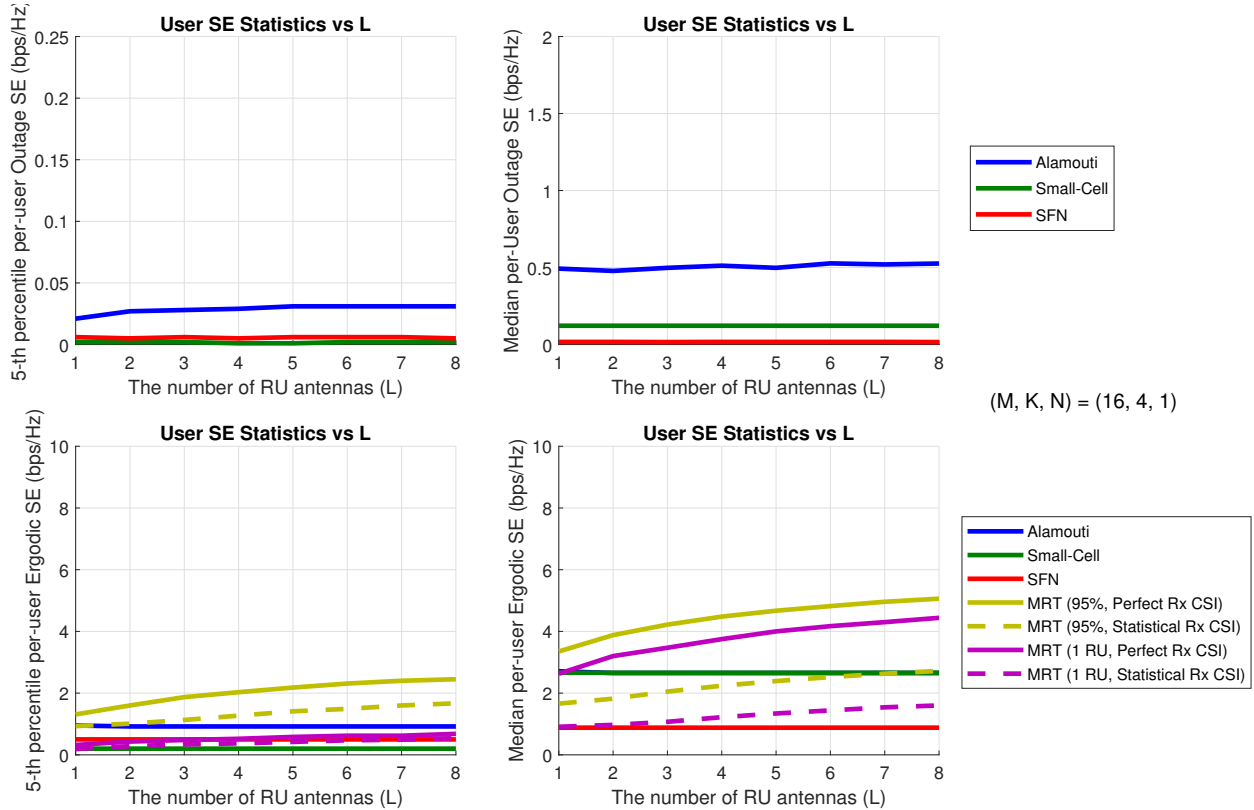}
\caption{$5$-th percentile and median SEs for various $L$ values ($M=16, K=4, N=1$).}
\label{fig10}	
\end{figure}

As shown in Fig. 10, the effect of the number of RU antennas to the performance of orthogonal coding is very limited. For $L>1$, MRT (1 RU) has better median ergodic SEs than orthogonal coding. This can be explained by the \textit{beamforming effect}. MRT can increase the desired signal strength applying beamforming through multiple RU antennas whereas methods without Tx CSI transmit the signal omni-directionally. Nevertheless, for all $L$ values, Alamouti has better $5$-th percentile ergodic SEs than MRT (1 RU), small-cell and SFN.

\newpage

\begin{figure}[ht]
	\centering
	\includegraphics[width=\linewidth]{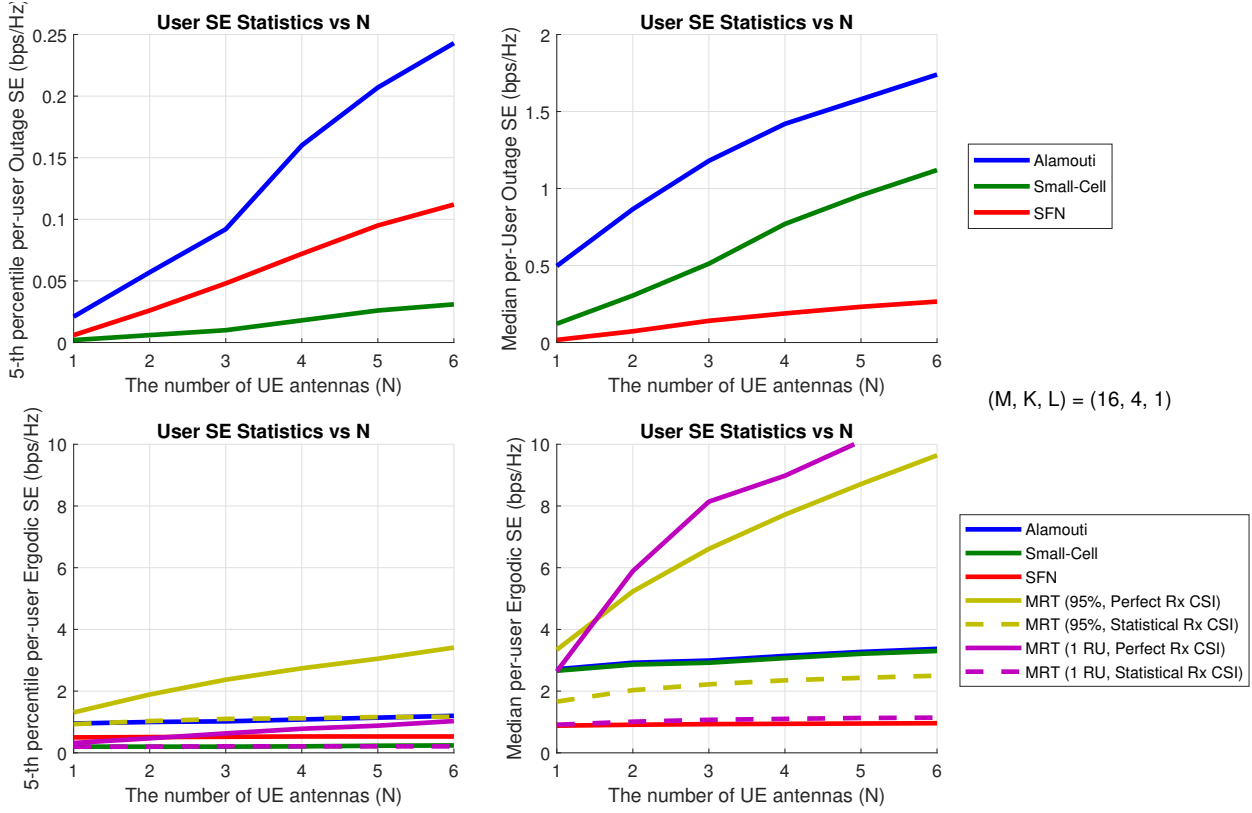}
\caption{$5$-th percentile and median SEs for various $N$ values ($M=16, K=4, L=1$).}
\label{fig11}	
\end{figure}

When we consider the effect of $N$, the number of antennas of UEs, given in Fig. 11, we observe a significant improvement in both outage and ergodic SEs for all methods. For $N>1$, MRT (1 RU) has much better median ergodic rates than orthogonal coding. This can be explained by \textit{multi-layer transmission effect}. Although multiple symbols are transmitted within a code period of the orthogonal code, as these symbols are transmitted at different time/frequency points, there is no any true multi-layer transmission. On the other hand, by means of the precoding at Tx side, MRT can transmit multiple independent symbols to UEs with multiple antennas. As a result, as $N$ increases, the median ergodic SE for methods with Tx and Rx CSI becomes much larger than those of methods without Tx CSI. The performances of the methods with statistical CSI at Rx side suffer from the unknown channel variations and hence their performances are not satisfactory. As a final remark, we observe that despite the multi-layer transmission effect, by means of optimized clustering, orthogonal coding has better $5$-th percentile per-user ergodic SEs than several methods with Tx CSI.

\newpage

\begin{figure}[ht]	
	\centering
	\begin{subfigure}{0.5\linewidth}
		\centering
		\includegraphics[width=0.9\linewidth]{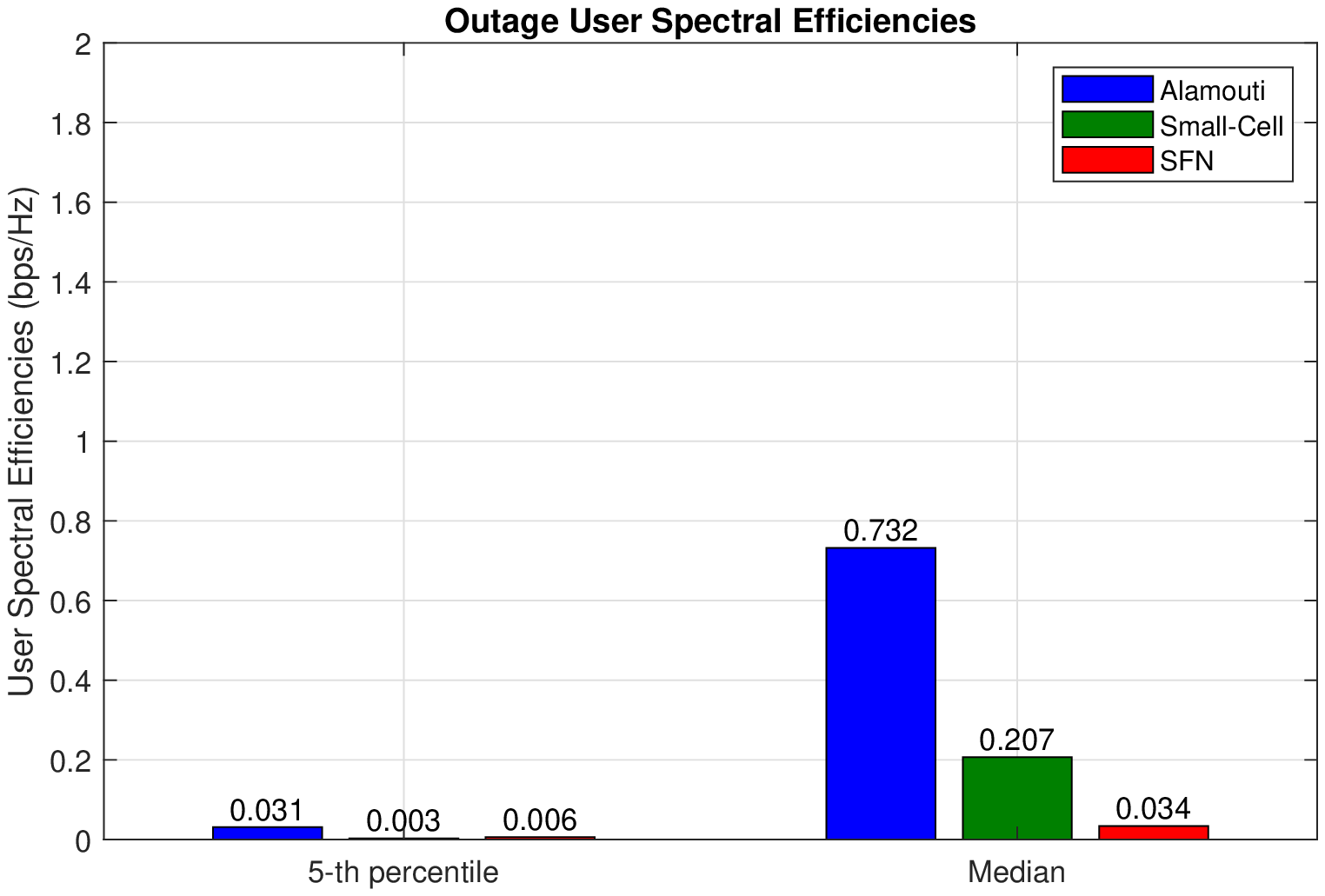}
		\caption{$5$-th percentile and median of outage SE}
	\end{subfigure}%
	\begin{subfigure}{0.5\linewidth}
		\centering
		\includegraphics[width=0.9\linewidth]{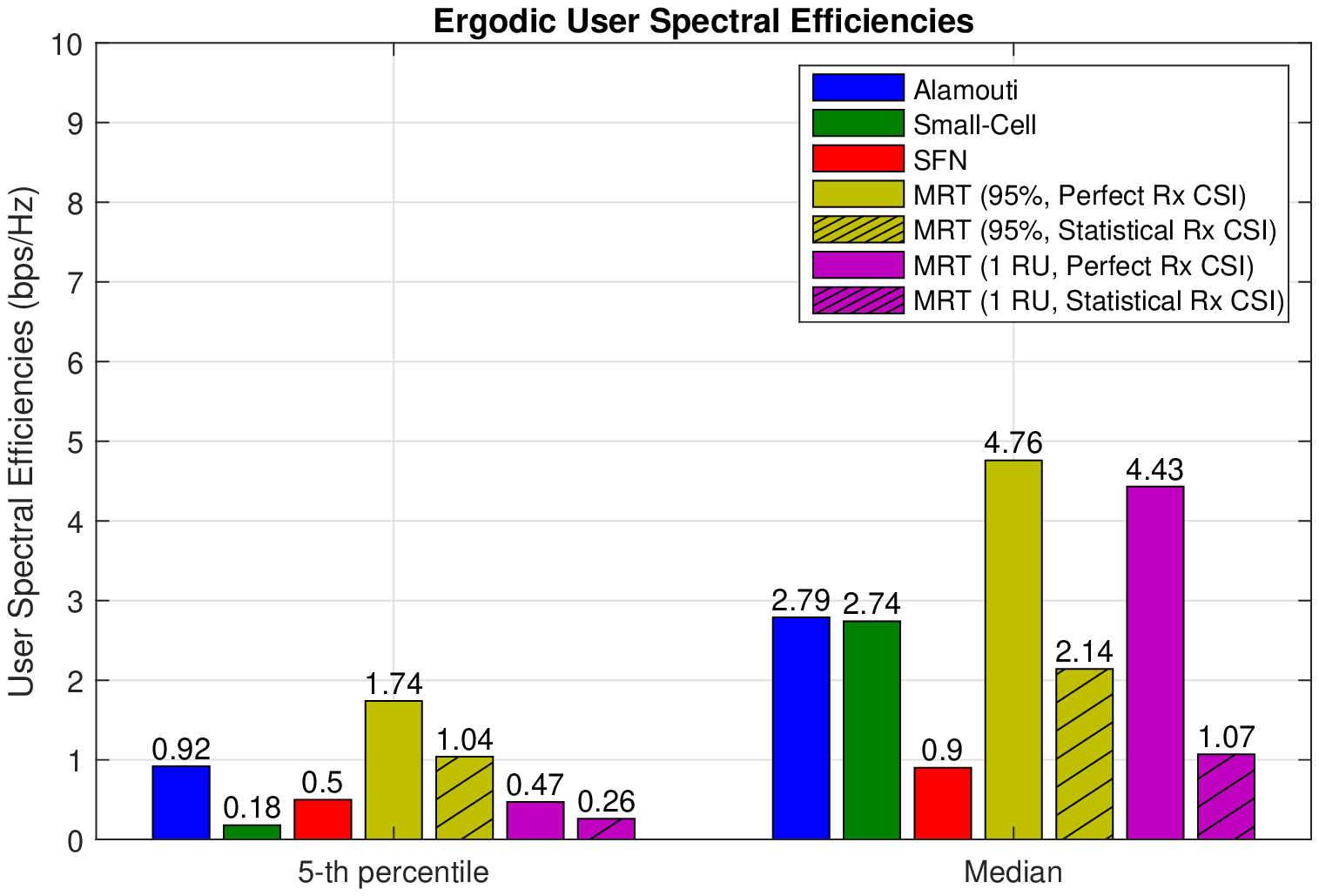}
		\caption{$5$-th percentile and median of ergodic SE}
	\end{subfigure}
\caption{$5$-th percentile and median SEs for all simulations.}
\label{fig12}	
\end{figure}

In Fig. 12, we present the results considering all simulations performed. We conclude that orthogonal coding has $3.5$ times better median and $10$ times better $5$-th percentile outage SEs than small-cell. Furthermore, by optimizing clustering, we obtain better $5$-th percentile ergodic SEs than MRT (1 RU), small-cell and SFN. In terms of median ergodic rates, which can be dramatically enhanced by beamforming and multi-layer transmission effects obtained by Tx CSI, MRT with perfect Rx CSI outperforms orthogonal coding. On the other hand, MRT with statistical Rx CSI has lower median ergodic rates than orthogonal coding due to insufficient CSI at Rx. 

\section{Conclusion}

In this study, we have investigated the potential benefits of Alamouti-like orthogonal codes in D-MIMO networks. We know that there may be some cases where accurate channel estimation for downlink channels may not be performed at RU side due to problems related to high mobility, lack of uplink/downlink reciprocity, and pilot contamination. In such cases, as a robust transmission scheme, the network can switch to orthogonal coding to increase diversity at UE side without using the instantaneous channel estimates. The results show that the proposed scheme has significant advantages about SE performance. It significantly outperforms other baseline methods small-cell and SFN in all cases. The results also reveal that when RUs and UEs have a single antenna, orthogonal coding together with an optimized clustering is better than MRT (1 RU) and has a close performance to MRT ($95\%$). We note that cluster formation is important to get benefits of orthogonal coding. As a future work one can investigate machine learning for clustering to further improve the performance.

\end{document}